\documentclass[12pt,preprint]{aastex}

\shorttitle{Young Triple V807 Tau}
\shortauthors{Schaefer et al.}

\begin{document}

%% LaTeX will automatically break titles if they run longer than
%% one line. However, you may use \\ to force a line break if
%% you desire.

\title{Orbit and Stellar Properties of the Young Triple V807 Tau}

%% Use \author, \affil, and the \and command to format
%% author and affiliation information.
%% Note that \email has replaced the old \authoremail command
%% from AASTeX v4.0. You can use \email to mark an email address
%% anywhere in the paper, not just in the front matter.
%% As in the title, use \\ to force line breaks.

\author{G. H. Schaefer\altaffilmark{1}, L. Prato\altaffilmark{2}, M. Simon\altaffilmark{3}, and R. T. Zavala\altaffilmark{4}}

\altaffiltext{1}{The CHARA Array of Georgia State University, Mount Wilson Observatory, Mount Wilson, CA 91023, U.S.A. (schaefer@chara-array.org)}
\altaffiltext{2}{Lowell Observatory, 1400 West Mars Hill Road, Flagstaff, AZ 86001, USA}
\altaffiltext{3}{Department of Physics and Astronomy, Stony Brook University, Stony Brook, NY 11794, USA}
\altaffiltext{4}{United State Naval Observatory, Flagstaff Station, 10391 W. Naval Obs. Rd., Flagstaff, AZ 86001, USA}

%% Notice that each of these authors has alternate affiliations, which
%% are identified by the \altaffilmark after each name.  Specify alternate
%% affiliation information with \altaffiltext, with one command per each
%% affiliation.

%% Mark off your abstract in the ``abstract'' environment. In the manuscript
%% style, abstract will output a Received/Accepted line after the
%% title and affiliation information. No date will appear since the author
%% does not have this information. The dates will be filled in by the
%% editorial office after submission.

\begin{abstract}

We present new orbital measurements of the pre-main sequence triple system, V807 Tau, using adaptive optics imaging at the Keck Observatory.  We computed an orbit for the close pair, V807 Tau Ba$-$Bb, with a period of 12.312 $\pm$ 0.058 years and a semi-major axis of 38.59 $\pm$ 0.16 mas.  By modeling the center of mass motion of the components in the close pair relative to the wide component, V807 Tau A, we measured a mass ratio of 0.843 $\pm$ 0.050 for Bb/Ba.  Combined with the total mass from the relative orbit, we derived individual masses of $M_{\rm Ba} =  0.564 \pm 0.018~(\frac{d}{\rm 140 pc})^3$ M$_\odot$ and  $M_{\rm Bb} =  0.476 \pm 0.017~(\frac{d}{\rm 140 pc})^3$ M$_\odot$ at an average distance of 140 pc to the Taurus star forming region.  We computed spectral energy distributions to determine the luminosities of the three components.  We also measured their spectral types, effective temperatures, and rotational velocities based on spatially resolved spectra obtained at the Keck Observatory.  If the rotational axes are aligned, then the projected rotational velocities imply that V807 Tau Ba and Bb are rotating much faster than V807 Tau A.  The uncertainty in the stellar effective temperatures and distance to the system currently limit the comparison of our dynamical mass measurements with predictions based on evolutionary tracks for pre-main sequence stars.  We also report preliminary results from a program to map the 3.6\,cm radio emission from V807 Tau using the Very Long Baseline Array.  With continued monitoring, these observations will provide a precise parallax for placing the dynamical masses on an absolute scale.

%The uncertainty in the distance contributes an additional systematic uncertainty of $\pm$ 0.24 M$_\odot$ to the masses.  

%spatially resolved high resolution spectra
% With continued monitoring in the future, the VLBA observations...

\end{abstract}

%% Keywords should appear after the \end{abstract} command. The uncommented
%% example has been keyed in ApJ style. See the instructions to authors
%% for the journal to which you are submitting your paper to determine
%% what keyword punctuation is appropriate.

\keywords{binaries: visual --- stars: pre-main sequence --- stars: fundamental parameters --- stars: individual (V807 Tau)}

%% From the front matter, we move on to the body of the paper.
%% In the first two sections, notice the use of the natbib \citep
%% and \citet commands to identify citations.  The citations are
%% tied to the reference list via symbolic KEYs. The KEY corresponds
%% to the KEY in the \bibitem in the reference list below. We have
%% chosen the first three characters of the first author's name plus
%% the last two numeral of the year of publication as our KEY for
%% each reference.

%% Authors who wish to have the most important objects in their paper
%% linked in the electronic edition to a data center may do so by tagging
%% their objects with \objectname{} or \object{}.  Each macro takes the
%% object name as its required argument. The optional, square-bracket 
%% argument should be used in cases where the data center identification
%% differs from what is to be printed in the paper.  The text appearing 
%% in curly braces is what will appear in print in the published paper. 
%% If the object name is recognized by the data centers, it will be linked
%% in the electronic edition to the object data available at the data centers  
%%
%% Note that for sources with brackets in their names, e.g. [WEG2004] 14h-090,
%% the brackets must be escaped with backslashes when used in the first
%% square-bracket argument, for instance, \object[\[WEG2004\] 14h-090]{90}).
%%  Otherwise, LaTeX will issue an error. 

\setcounter{footnote}{4}

\section{Introduction}

V807 Tau (Elias 12, HBC 404) is a pre-main sequence (PMS) hierarchical triple system located in the Taurus star forming region.   The system was resolved into two wide components at a separation of $\sim$ 300 mas by \citet{ghez93,ghez95}, \citet{leinert93}, and \citet{simon96}.  \citet{white01} and \citet{hartigan03} identified the primary (V807 Tau A, Elias 12 S) as a K5$-$K7 classical T Tauri star showing signs of an accretion disk, while the secondary (V807 Tau B, Elias 12 N) was identified as a M2$-$M3 weak-lined T Tauri star with no evidence for accretion.  \citet{simon95} resolved the secondary into two close components (V807 Tau Ba$-$Bb, Elias 12 Na$-$Nb) during a lunar occultation.  This close pair is separated by $\sim$ 40 mas.  We have been mapping the orbital motion of V807 Tau as a triple system beginning in 1999 using the Fine Guidance Sensors (FGS) onboard the {\it Hubble Space Telescope} and later with adaptive optics imaging at the Keck Observatory \citep{schaefer03,schaefer06}.

Mapping the orbital motion of a binary star provides a fundamental way to measure the dynamical masses of the component stars.  This can be accomplished by observing eclipsing double-lined spectroscopic binaries where the radial velocity variations and timing of the eclipse light curve yield the masses and radii of the component stars \citep[e.g.][]{andersen91}.  Alternatively, spatially resolving the visual orbit of a double-lined spectroscopic binary (SB2) yields the component stellar masses and distance to the system.  At the distances to the nearest star forming regions \citep[140 pc for the Taurus star forming region;][]{kenyon94}, resolving short period PMS SB2s becomes challenging and requires high angular resolution techniques \citep[e.g.][]{steffen01,boden05,boden07,schaefer08}.  With an orbital period of $\sim$ 12 years (Sect.~\ref{sect.orbit}), the velocity separation of the close binary components in V807 Tau are no more than a few km\,s$^{-1}$.  Attempts to measure the radial velocity variations are further complicated because the spectra of V807 Tau Ba$-$Bb are rotationally broadened to velocity widths of 60$-$70 km\,s$^{-1}$ (Sect.~\ref{sect.nirspec}).  

Fortunately, the triple nature of V807 Tau provides an additional technique for determining the masses of the components in the close pair.  When combined with an estimate for the distance to the system, the relative visual orbit between V807 Tau Ba$-$Bb provides a measurement of their total mass.  Additionally, the wide primary V807 Tau A provides a reference position to measure the astrometric motion of V807 Tau Ba$-$Bb around their center of mass, providing a direct measurement of the mass ratio between the close components.  Together, these measurements give the individual masses of the components in V807 Tau Ba$-$Bb.  A similar approach was followed to determine the masses for the components in the close pair in the T Tau triple system \citep{duchene06}.

In this paper we present the results of a multi-wavelength campaign to study the V807 Tau triple system.  We used infrared adaptive optics imaging at the Keck Observatory to map the astrometric motion of the close pair and present the orbital solution and component masses in Sect.~\ref{sect.orbit}.  We also obtained spectra of the three components on several epochs using a high spectral resolution spectrometer operating with the adaptive optics system at the Keck Observatory to separate out the contribution of each component.  Although we could not measure radial velocities precisely enough to determine a spectroscopic orbit, we discuss the spectral types and rotational properties of the stars within the system in Sections~\ref{sect.nirspec} and \ref{sect.rot}.  Using estimates of the effective temperature and luminosity computed by fitting the spectral energy distribution for each component, we compare our dynamical measurements with predicted masses and ages from PMS evolutionary tracks in Section~\ref{sect.tracks}.  We also report on the detection of 3.6\,cm radio emission measured with the Very Long Baseline Array in Sections~\ref{sect.vlba} and \ref{sect.ir_radio}.  The location of the radio emission is consistent with the infrared locations of the components in the close pair.  With continued observations, the radio measurements could provide an independent measurement of the parallax of V807 Tau.  

\section{Observations and Data Reduction}
\label{sect.obs}

\subsection{Adaptive Optics Imaging}

We obtained adaptive optics (AO) images of V807 Tau using the near-infrared camera NIRC2 \citep{wizinowich00} on the 10-m Keck II Telescope at the W. M. Keck Observatory.  The NIRC2 detector is a 1024$\times$1024 Aladdin-3 InSb array.  The images were taken with the narrow-field camera which has a plate scale of 9.952$\pm$0.002 mas~pixel$^{-1}$ \citep{yelda10}.  During each epoch, we obtained 10$-$20 sets of dithered images, using a three-point dither pattern with a dither offset of $\sim 2 ''$.  The primary star V807 Tau A was used as the AO guide star.  

%\footnote{http://www2.keck.hawaii.edu/inst/nirc2/genspecs.html}

\begin{figure}
\begin{center}
  \scalebox{0.7}{\includegraphics{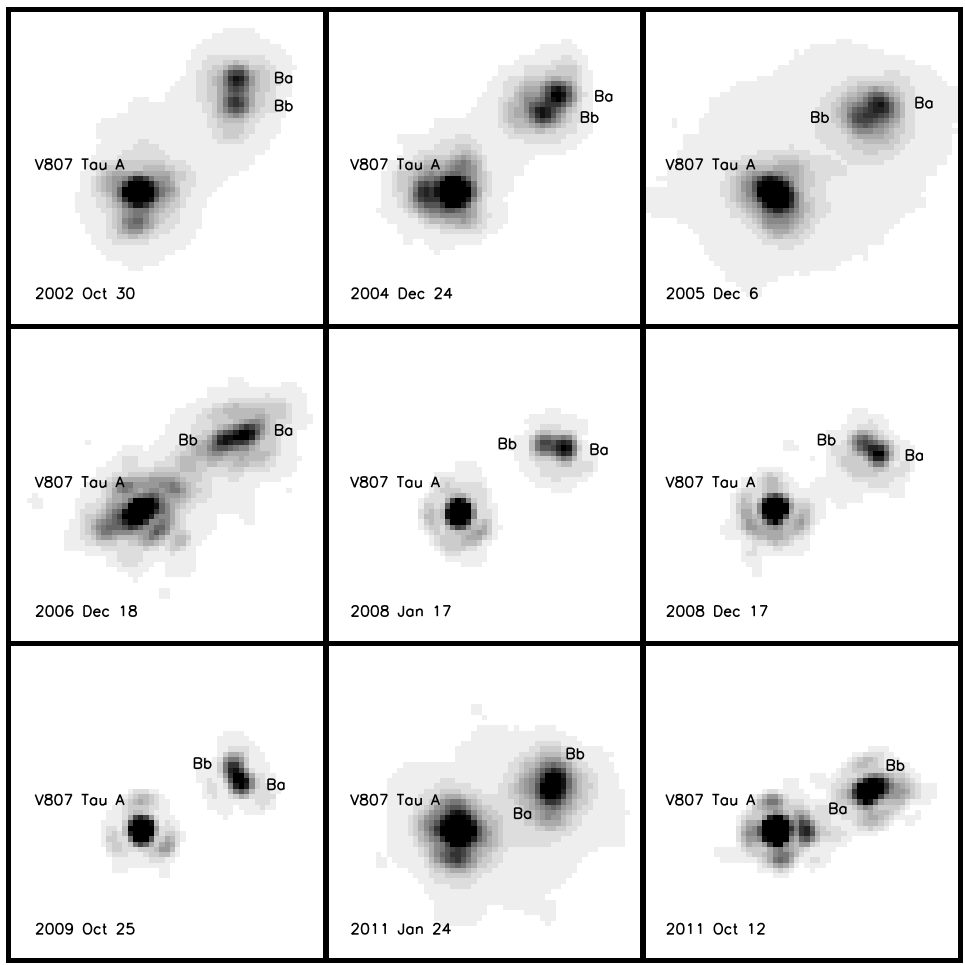}}
\end{center}
  \caption{Keck NIRC2 adaptive optics images in the $H$-band of V807 Tau obtained in 2002$-$2011.  Each panel is $0.''6$ wide, with north pointing up and east to the left.  We used the wide component, V807 Tau A, as a PSF reference to model the components in the close pair.  The shape of the AO-corrected PSF changes over time; this is seen most prominently in the location of the bright spots in the diffraction ring or an elongation of the core.  However, for any given image, the shape of the PSF of V807 Tau A provides an excellent model of the PSFs of Ba and Bb during the same instant of time, allowing us to measure high precision positions and flux ratios of the close components.}
\label{fig.images}
\end{figure}

Table~\ref{tab.ao-obs} lists the AO observations of V807 Tau taken since \citet{schaefer06}.  The table lists the filters, AO correction rates, exposure time, number of co-added exposures per image, and total number of images obtained.  The images were flatfielded using dark-subtracted, median filtered dome flat fields obtained on the nights of observation.  Pairs of dithered exposures were subtracted to remove the sky background.  Figure~\ref{fig.images} shows co-added AO images for V807 Tau from 2002$-$2011.

The wide primary V807 Tau A is sufficiently bright and close, within the isoplanatic patch, to V807 Tau B that it can be used as a simultaneous point-spread function (PSF) for measuring the separation and flux ratio of the close pair.  We extracted subarrays from the images centered on the close pair, with widths of $0\farcs12 - 0\farcs16$.  We used V807 Tau A as a PSF to construct models of the close pair while searching through a grid of separations and flux ratios.  We used the IDL INTERPOLATE procedure to shift the PSF by sub-pixel intervals using cubic convolution interpolation.  To remove any excess light of the primary star from the extracted sub-array of the close pair, we followed a least-squares approach to fit the background as a slanting plane.  We determined the position and flux ratio of the companions in the close binary from where the $\chi^2$ between the model and the data reached a minimum.  After referencing the location of the close pair to V807 Tau A in the full array, we corrected the positions for geometric distortion by using the distortion solution computed by \citet{yelda10}.  We subtracted 0.252$^\circ$$\pm$0.009$^\circ$ from the measured position angles to correct for the orientation of the NIRC2 camera relative to north.  We determined uncertainties in the positions and flux ratios by analyzing multiple images individually and computing the standard deviation.

%After referencing the location of the close pair to V807 Tau A in the full array, we corrected the positions for geometric distortion by using the IDL routine nirc2dewarp.pro written by P. B. Cameron\footnote{http://www2.keck.hawaii.edu/inst/nirc2/forReDoc/post\_observing/dewarp/nirc2dewarp\_positions.pro}.

Table~\ref{tab.sepPA} lists the separation and position angle measured for V807 Tau A$-$Ba, A$-$Bb, and Ba$-$Bb.  The position angle is measured east of north and is referenced to the date of observation.  For completeness, we also include previously published positions from \citet{schaefer06}.  We have revised the uncertainties for the close pair from the previously published values because we now use the standard deviation determined directly from the relative separation and position angle of the close pair rather than propagating the uncertainties derived from the widely spaced separations of A$-$Ba and A$-$Bb.  We also corrected the previously published positions for geometric distortions in the camera.

The cumulative listing of infrared flux ratios measured for the V807 Tau triple system from the adaptive optics observations is presented in Table~\ref{tab.aoflux} which lists the flux ratios of Ba relative to A, Bb relative to A, and Bb relative to Ba.  In the lower portion of the table we list weighted mean values for the flux ratios in the $J$, $H$, $K'$, and $L'$ bands.  We also used the photomultiplier (PMT) counts from the Fine Guidance Sensor observations (using the F583W filter) published in \citet{schaefer06} to estimate the V-band flux ratios.  In Table~\ref{tab.compmag}, we convert the flux ratios to component magnitudes by using published photometry obtained for the unresolved system.  The top portion of the table lists the published photometry for the total flux of V807 Tau \citep{kharchenko01,kharchenko09,cutri03} as well as magnitudes for the separated northern and southern components \citep{white01}.  In the lower portion of the table we use these total magnitudes along with the average flux ratios in Table~\ref{tab.aoflux} to compute the magnitude of each component.  We also use the FGS magnitude conversion \citep{holfeltz95,schaefer03} to convert directly the mean PMT counts into V-band magnitudes of the components \citep[using  $B-V$ $\sim$ 1.2;][]{white01}.

\subsection{High Resolution Spectroscopy}
\label{sect.nirspec}

To measure radial and rotational velocities and determine spectral types, we obtained spectra with NIRSPEC, the near-infrared, cryogenic, cross-dispersed echelle spectrograph \citep{mclean98, mclean00} on the Keck II telescope. Observations with NIRSPEC behind the AO system (NIRSPAO) separated the light of the V807 Tau components; additional unresolved spectra were taken with NIRSPEC alone.  We centered the observations at 1.555 $\mu$m and used the 2 pixel wide slit which yielded a spectral resolution $\sim 30,000$.  NIRSPAO provides a pixel scale of $0\farcs018$ in the cross-dispersion direction, about 10.6 times finer than the $0\farcs190$ scale for NIRSPEC.  Table~\ref{tab.nirspec_obs} summarizes the NIRSPEC and NIRSPAO observations.  The diffraction-limited resolution at 1.6 $\mu$m is $\sim 0\farcs040$.   Columns 2 and 3 list the orbital phase and separation of V807 Tau Ba$-$Bb during the times of the NIRSPAO observations predicted from the visual and astrometric orbital parameters (see Sect~\ref{sect.orbit}).  For the 2001 and 2009 AO observations, we aligned the slit along the position angle of the Ba$-$Bb binary; we also observed the A component separately in 2001.  In 2003, 2006, and 2010, the slit was aligned along the A$-$B axis which was fortuitously similar to the Ba$-$Bb axis in 2003 and 2006.  In 2010 the close binary was separated by less than 30 mas and thus unresolvable.  Successful separation of the Ba$-$Bb components was possible only for the 2001 and 2003 data.  Nodding along the slit between sequential exposures by about half the slit length enabled removal of the sky background.  Exposure times ranged from 60 to 360 seconds; because of the reduced throughput behind the AO system, NIRSPAO spectra required longer integrations.

The spectra were reduced and extracted using the REDSPEC software package\footnote{http://www2.keck.hawaii.edu/inst/nirspec/redspec.html}.  The software performs background subtraction, flat-fielding, spatial and spectral rectification, and subsequent extraction of the spectra.  The spatial and spectral rectification removes curvature distortions in order to produce straight spectral traces.  For non-AO observations the dispersion solution and wavelength calibration were computed with respect to OH night-sky lines \citep{rousselot00}.  As a result of the reimaged optics and reduced throughput in the AO mode, the OH night sky lines in NIRSPAO spectra are too faint to be used for wavelength calibration and the internal Ne, Ar, Xe, and Kr arc lamps were used instead.  For the purpose of measuring radial velocities, only the spectra in order 49 were extracted and used in the subsequent analysis.  In addition to atomic lines, this spectral region (Figure 2) contains molecular lines, most notably OH and H$_2$O, suitable for characterizing M and late-K spectral types, and is free of terrestrial absorption features \citep{prato98, prato02, mazeh02}.  
  
On most NIRSPAO runs, resolved spectra of V807 Tau A and B were obtained, however, V807 Tau Ba and Bb were always blended because their separations were close to or below the diffraction limit.  Nonetheless, when the components were at favorable separation and the AO correction particularly good, the $0\farcs018$ pixel$^{-1}$ scale of the detector in the cross-dispersion direction enabled us to extract individual spectra for Ba and Bb by fitting two Gaussians to the spatial profile of the data at each pixel, and thus wavelength, across order 49.  We were able to extract unique spectra of Ba and Bb on 2001 Jan 8 and 2003 Dec 13 but were unable to separate the spectra from the later NIRSPAO runs when the components were at smaller separations.  The unresolved spectra are designated Ba+Bb in Table~\ref{tab.nirspec_obs}.

We used our suite of NIRSPEC spectral templates \citep{prato02, bender05} to identify by cross-correlation the template that best represented the spectrum of V807 Tau A and to measure its heliocentric radial velocity and rotational broadening.  We determined the best fit by varying the spectral template and the applied rotational broadening to produce the highest cross-correlation coefficient; the radial velocity is given by the location of the peak.  HR 8086, a K7V spectral type star, gave the best fit for V807 Tau A; the  measured heliocentric radial velocities and rotational broadening are listed in Table~\ref{tab.nirspec_vel}.   The average value and standard deviation of the radial velocity, $15.79 \pm 0.65$ km~s$^{-1}$,  is a good approximation to the systemic velocity of the V807 Tau triple\footnote{At the $\sim 140$ pc distance of the Taurus star forming region  the projected orbital separation of the  $0.''3$ V807 Tau A$-$B triple is about 43 AU.  Assuming a circular orbit and total mass 1.5 M$_\odot$, the period is about 230 yrs, much greater than the orbital period of the Ba$-$Bb binary.}.  The small downward trend in radial velocity at a rate of $-0.15 \pm 0.04$ km~s$^{-1}$ per year is consistent with the amount of orbital motion that would be expected between the wide A-B pair.
The broadening of the spectral lines indicates that V807 Tau A is rotating with a projected velocity of $\sim 15 \pm 3$ km~s$^{-1}$.

We applied the same analysis to the angularly resolved spectra of V807 Tau Ba and Bb.   We obtained the best fit to the Ba spectrum with the template of Gl 752A, an M2V spectral type star, rotationally broadened to 65$-$70 km s$^{-1}$. The result for Bb is similar, the spectra are best fit by Gl 436 (M2.5V) and Gl 752A  rotationally broadened to 60$-$70 km s$^{-1}$.  The difference in spectral types of the two fits is not significant because the precision of spectral type determination is no better than $\pm$ one subtype.  The spectral types are consistent with the determinations of the unresolved Ba$-$Bb binary from broadband photometry \citep{white01} and the visible light spectrum \citep{hartigan03}.   We were unable to measure consistent radial velocities of Ba and Bb using stellar templates, or consistent velocity differences between the radial velocities of Ba and Bb by using one spectrum as the template for the other.  We estimate the uncertainty of these measurements as $\ge \pm 5$ km s$^{-1}$.   We were also unable to measure consistent radial velocities of the unresolved Ba $+$ Bb spectra  by two-dimensional cross-correlation with stellar templates.  Figure~\ref{fig.predRV} shows predicted radial velocities versus phase of Ba and Bb and illustrates the difficulty in measuring their velocities.  The predicted velocities were calculated using the orbital parameters and mass ratio $M_{\rm Bb}/M_{\rm Ba}$ = 0.843 in Section~\ref{sect.orbit}, systemic velocity, and 140 pc distance to the Taurus star forming region.  The predicted velocity differences are never greater than $\sim 7.7$ km s$^{-1}$.   It was not possible to distinguish such small differences in radial velocities of Ba and Bb given the significant broadening of their spectra. 

The observed templates used for the cross-correlation analysis are main sequence field objects while the components in V807 Tau are young stars still contracting to the main sequence.  Hence, the PMS stars will have smaller surface gravities by about 0.3 to 0.5 in $\log{g}$.  Metal lines are weakly dependent on $\log{g}$ \citep{blake07,schlieder12}; the line centers remain the same while the equivalent width of most lines decreases with increasing $\log{g}$.  Therefore, the gravity dependence has no effect on our measurement of the radial velocities which are dominated by the location of the lines \citep[e.g.][]{simon04,rosero11} and only a negligible effect on the rotational broadening for lines as broad as those of V807 Tau Ba and Bb.

\subsection{VLBA Radio Interferometry}
\label{sect.vlba}

Radio emission from V807 Tau was first detected by \citet{oneal90} using the VLA.  During two out of three epochs in 1986-1987, they measured a flux density of 0.7$-$1.2 mJy at 6\,cm.  We began a program to measure the radio emission at high resolution using the Very Long Baseline Array (VLBA) in 2007$-$2009.  Our goals were to determine where in the system the radio emission comes from and to map the astrometric position over time to determine a parallax for the system \citep[e.g.][]{lestrade99,loinard05,loinard08,torres07,torres09,torres12}.

The VLBA \citep{napier94,napier95} consists of ten 25-meter antennas spread across the United States territory from Saint Croix, VI to Mauna Kea, HI with baselines ranging from 236 to 8611 km.   The signals are combined interferometrically by a correlator in Socorro, NM.  We obtained six epochs of phase-referenced observations on V807 Tau at 3.6\,cm (8.42 GHz) during 2007$-$2009.  Each observation lasted a total of 10 hours, with $\sim$~4 hours spent on V807 Tau.  During each observation, we observed in cycles spending 2 minutes on V807 Tau and 1 minute on the primary phase reference calibrator J0426+2327=0423+233 located 1.6$^\circ$ away ($F_\nu =$ 0.14 Jy at 3.6cm)\footnote{http://www.vlba.nrao.edu/astro/calib/vlbaCalib.txt}.  Every 30 minutes we observed 1 minute integrations on the phase reference check sources J0429+2724=0426+273 and J0450+2249=0447+227 located 3.3$^\circ$ and 4.3$^\circ$ away, respectively ($F_\nu$ = 0.32 and 0.12 Jy).  The location of these three phase sources on the sky provides good coverage around V807 Tau, as shown in Figure~\ref{fig.vlbapos}.  During the first exploratory observation, we used only J0426+2327 and J0429+2724 and spent 3 minutes on source and 2 minutes on the phase reference calibrators.  In all of the epochs, we observed the fringe finder 3C\,84 twice during each night.  We also observed three sets of 6$-$7 calibration sources covering a range of elevations spaced apart in time by about 4 hours.  These sources were observed using a wide bandwidth frequency setup to compute multi-band delays and correct for errors in the troposphere model.

We reduced and calibrated the VLBA data using the Astronomical Image Processing System \citep[AIPS;][]{vanmoorsel96} developed by NRAO following the standard procedures for reducing VLBA data.  This includes applying ionospheric corrections, updating Earth orientation parameters, amplitude calibration, correction to phases for the parallactic angle of the antennas, and removal of instrumental delay residuals.  We corrected for errors in the tropospheric models by using the DELZN procedure \citep{mioduszewski09} to model the multi-band delay on the wide-band calibrator sets.  We performed a global fringe fit to remove global frequency and time dependent errors.  We used the primary and secondary phase calibrators to measure and correct for the phase gradient on the sky using the ATMCA routine \citep{fomalont05}.  We produced cleaned images using the difmap routine\footnote{ftp://ftp.astro.caltech.edu/pub/difmap/difmap.html} \citep{sheperd97} with natural weighting (equal weighting of all data points) to provide higher sensitivity on weak sources.  Reading the image back into AIPS, we used the JMFIT task to solve for the position, integrated intensity, and angular size of the source by fitting a two-dimensional Gaussian model to the data.  The results from the fit are presented in Table~\ref{tab.vlba} which gives the RA and DEC coordinates, integrated intensity, major and minor axes (FWHM), and position angle of the Gaussian model.  We detected compact radio emission in three of the six epochs of VLBA observations.  On 2007 Mar 10 and 2008 Aug 28 we detected a single source and on 2008 Mar 19 we detected two sources.   The integrated intensity ranges between 0.52 to 8.60 mJy.  The non-detections during the remaining three epochs were likely caused by intrinsic variability of the source.  By imaging the phase reference and check sources during each of the VLBA epochs, we determined that our measured source coordinates are consistent with the catalog positions of the calibrators to better than 0.15~mas.  Figure~\ref{fig.vlba_map} shows the cleaned images of V807 Tau during the three epochs with detected emission.  The position of the radio emission is consistent with the location of the components in the close pair, V807 Tau Ba and Bb (see Sect.~\ref{sect.ir_radio} for more details).

% at a separation and position angle consistent with the expected location of V807 Tau Ba and Bb in the infrared

The radio emission from V807 Tau is most likely non-thermal and 
thus is expected to be polarized. While the phase-referenced observations 
were not planned with polarimetry in mind, we performed a polarization
calibration in order to learn what we could from detections or limits 
on the polarized flux.  
We solved for the right-left (R$-$L) phase difference \citep{cotton93} using the AIPS task RLDLY and used our weakly polarized fringe-finder 3C\,84 to determine the R$-$L delay.  We solved for the instrumental polarization (D-terms) using the AIPS task LPCAL \citep{leppanen95} applied to the phase reference source J0426+2327. To examine the success of our polarization calibration 
we imaged 3C\,84 in Stokes Q and U and made a polarized intensity image. 
We found the southern lobe of 3C\,84 to be weakly polarized. Fortuitously, 
calibrated polarization images of 3C\,84 were 
available via the publicly accessible MOJAVE archive \citep{lister09}. 
3C\,84 was observed on 2008 Aug 25 and the southern lobe was weakly 
polarized at a location coincident with our detection. Satisfied that the 
polarization calibration worked well enough we imaged V807 Tau in 
Stokes Q, U and V using the 2008 Aug 28 data. This date had a sufficient 
Stokes I flux that even a small fractional polarization might be detected.
We found no detectable polarized flux at this epoch and set 3\,$\sigma$ 
limits on the fractional linear polarization of $<$ 11\% and on the 
fractional circular polarization of $<$ 8\%.

\section{Analysis}

\subsection{Orbital Solution}
\label{sect.orbit}

Figure~\ref{fig.orbit} shows the relative motion observed for the close pair based on the measurements listed in Table~\ref{tab.sepPA}, where the position of V807 Tau Bb is plotted relative to V807 Tau Ba.  Over 10 years of FGS and AO observations, the close pair has moved through a complete orbit.  We computed an orbit for the close pair using a standard Newton Raphson technique.  Table~\ref{tab.orbit} lists the best fitting orbital parameters for the period $P$, time of periastron $T$, eccentricity $e$, semi-major axis $a$ (in mas), inclination $i$, position angle of the line of nodes $\Omega$, and the longitude of periastron $\omega$.  The $\chi^2$ and reduced $\chi^2_\nu$ of the fit are also listed in the table.  The uncertainties were estimated from the diagonal elements of the covariance matrix.    The orbit, with a period of 12.3 years and a semi-major axis of 38.6 mas, is overplotted in Figure~\ref{fig.orbit}.  At an average distance to the Taurus star forming region of 140 pc \citep{kenyon94}, Kepler's Third Law yields a total mass of $M_{\rm tot} = 1.040 \pm 0.016~(\frac{d}{\rm 140 pc})^3$~M$_\odot$.

In \citet{schaefer06} we presented a variety of possible orbit fits for V807 Tau Ba$-$Bb based on a limited number of early FGS, AO, and lunar occultation measurements (we refer to the system as ``Elias 12 Na-Nb'' in the previous paper).  Based on the preliminary analysis of orbits that fit the available data at the time, we originally estimated that the period of V807 Tau Ba$-$Bb ranged from 9$-$12 years.  This is quite close to the actual period of $12.312  \pm 0.058$ years that we measure with improved accuracy and precision because the orbital motion now covers a full orbital period.  Moreover, the total mass determined from the statistical analysis of orbits that fit the data presented in \citet[][1.13$^{+0.36}_{-0.09}$~M$_\odot$]{schaefer06} agrees within 1\,$\sigma$ with our results, confirming our earlier conclusion that it is possible to derive a reasonable estimate of the total mass in a binary with limited orbital coverage because of the tight correlation between $P$ and $a$.  The total mass we derived in \citet{schaefer06} when we added the lunar occultation observation from \citet{simon95} to improve the time coverage of the limited set of orbital measurements ($1.026 \pm 0.068$~M$_\odot$) also agrees quite well with our more precise current value.  We do not include the lunar occultation within our current solution because it represents only a one dimensional measurement of the projected separation along the direction of the occultation and we currently have sufficient orbital coverage to measure the period reliably without including it.  Nonetheless, the projected separation measured in the 1992 occultation agrees within 5.3~mas (2.3\,$\sigma$) with the separation predicted from our orbit fit.  Including the lunar occultation in the fit only marginally changes the orbital parameters within their 1\,$\sigma$ error bars, but it also raises the $\chi^2$ from 15.4 to 20.1 because of the 2.3\,$\sigma$ disagreement with the measured projected separation.

The wide component, V807 Tau A, is located within $\sim$250 mas of the close pair and lies within the adaptive optics field of view.  This allows us to map the astrometric motion of the close pair relative to V807 Tau A and model the motion of Ba and Bb around their center of mass.  Figure~\ref{fig.astrom} shows the motion of V807 Tau Ba and Bb as measured relative to V807 Tau A.  We derived a value for the mass ratio of the close pair ($M_{\rm Bb}/M_{\rm Ba}$) by modeling the center of mass motion of V807 Tau B relative to V807 Tau A.  To do this, we searched through a range of mass ratios and computed the expected location of the center of mass of V807 Tau Ba$-$Bb during each epoch based on the given mass ratio and the measured positions of Ba and Bb relative to V807 Tau A (Table~\ref{tab.sepPA}).  For each trial mass ratio, we then fit a representative orbit to the center of mass motion of B relative to A and selected the mass ratio that minimized the $\chi^2$ between the calculated position of the center of mass and the orbit fit (see Figure~\ref{fig.pos_cms}).  An incorrect mass ratio will produce residual reflex motion that cannot be fit by a simple Keplerian orbit.

We do not have enough coverage to determine a definitive orbit for the wide A-B pair, however, it is only necessary to find a representative orbit that adequately describes the orbital motion of the center of mass over the time-frame of our measurements.  In fact, there is a large range of orbits that can fit the limited data equally as well \citep[e.g. see Figures 3, 5, and 10 in][which show examples of how a family of possible orbits can fit a limited number of measurements for an incomplete orbit.]{schaefer06}.  To optimize the fit to the center of mass motion, we searched through a grid of possible orbits for each mass ratio explored.  The possible orbits were selected by randomly choosing $P$, $T$, $e$ over a range of values and performing a least-squares fit to optimize the remaining parameters following the method described in \citet{schaefer06} \citep[see also][]{hartkopf89}.  We searched through periods ranging from 100 to 500 years, eccentricities from 0 to 0.99, and allowed the time of periastron passage to vary within one orbital period.  For each mass ratio, we selected the ``best-fitting'' representative orbit that minimized $\chi^2$ between the calculated position of the center of mass and the orbit fit.  We then selected the mass ratio that produced the smallest $\chi^2$ as the best fit.  We determined the uncertainty in the mass ratio from the 1\,$\sigma$ confidence interval ($\Delta \chi^2 = 1$).  When combined with the total mass, the best fitting mass ratio of 0.843 $\pm$ 0.050 yields individual masses of $M_{\rm Ba} =  0.564 \pm 0.018~(\frac{d}{\rm 140 pc})^3$ M$_\odot$ and  $M_{\rm Bb} =  0.476 \pm 0.017~(\frac{d}{\rm 140 pc})^3 $ M$_\odot$.

The dynamical masses were computed using the average distance of 140~pc to the Taurus star forming region \citep{kenyon94} and scale as $(\frac{d}{\rm 140 pc})^3$.  Including the $\pm$~10~pc uncertainty in the average distance to the masses contributes an additional systematic uncertainty of $\pm$0.24 M$_\odot$ to the error budget.  Star formation in Taurus occurred in several clumps.  Several individual members in the region are known to have precise distances ranging from 128.5~pc for Hubble 4 and 161.2~pc for HP Tau/G2 \citep[][and reference therein]{torres09}.  The position on the sky of V807 Tau is near HP Tau/G2; if its distance is similar too, then the actual masses will be $(160/140)^3$ $\sim$ 1.5 times larger.  VLBA observations to measure the distance to V807 Tau (see Sect~\ref{sect.vlba} and \ref{sect.ir_radio}) are in progress.

\subsection{Spectral Energy Distributions}
\label{sect.sed}

We constructed spectral energy distributions (SED) for each of the three components, V807 Tau A, Ba, and Bb, using the magnitudes listed in Table~\ref{tab.compmag}.  We converted the observed magnitudes $m_\lambda$ to measured fluxes through 
\begin{equation}
F_{\lambda, \rm meas} = F_{\lambda_0} \times 10.0^{-0.4[m_\lambda - A_\lambda]}
\end{equation}
where $A_\lambda$ is the wavelength-dependent interstellar extinction assuming an R=3.1 reddening law \citep{cardelli89,odonnell94} and $F_{\lambda_0}$ is the zero-point flux density for a filter at wavelength $\lambda$.  We then fit low resolution spectral models computed by R.\ Kurucz and Castelli\footnote{http://kurucz.harvard.edu/grids.html}$^,$\footnote{http://wwwuser.oat.ts.astro.it/castelli/grids/gridp00k0odfnew/fp00k0tab.html} to the measured photometry.  We averaged the flux of the stellar models across the width of each of the observed filter bands.  The emergent flux from the star given by the models scales as the area subtended on the sky (or as the square of the stellar angular diameter).  Therefore, fitting the SED with stellar models is dependent on three parameters, the effective temperature $T_{\rm eff}$ of the stellar model, the angular diameter of the star $\theta$, and the interstellar extinction $A_V$ in the $V$-band.  The SED fits also provide an estimate of the bolometric flux $F_{\rm bol} = \onequarter \theta^2 \sigma T^4$ and the luminosity $L = 4 \pi d^2 F_{\rm bol}$.

We converted the spectral types determined from the NIRSPAO spectra to effective temperatures using the dwarf spectral type temperature relations used in \citet{kraus07} for K0$-$M0 and the T-Tauri temperature scale defined in \citet{luhman03} for M1$-$M7.  We fixed these temperatures when computing the SED fits and determined uncertainties by varying the model spectrum by one spectral step ($\sim$ 250 K) and calculating the range of variation in the derived parameters.  We know from the FGS photometry that V807 Tau A is variable \citep[see Fig.\ 15 in][]{schaefer06}; this is consistent with its classification as a classical T Tauri star.  To account for the variability potentially being propagated through to the magnitudes of the individual components, we applied a floor uncertainty of 0.1 mag to all of the magnitudes listed Table~\ref{tab.compmag}.  In Table~\ref{tab.sed}, we list the spectral type, adopted $T_{\rm eff}$, and the parameters derived from the SED fit including $A_V$, $\theta$, $F_{\rm bol}$, and $L$.  Figure~\ref{fig.sed} shows plots of the SED fits.  For V807 Tau A, we did not include the $K$ and $L$-band magnitudes in the fit because they are affected by the infrared excess from the circumstellar disk.  Therefore, the luminosity we calculate for V807 Tau A should mostly reflect the star, with little contamination from the disk.  The SED fits for V807 Tau Ba and Bb show that the infrared magnitudes are well matched by the spectral templates, confirming that there is no evidence for circumstellar disks around these weak-lined T Tauri stars \citep{white01,hartigan03}.  The luminosities we derive are in agreement with the estimates for the A$-$B components determined by \citet{white01}.  The presence of a disk can also be identified by a deredenned $K-L$ color in excess of 0.4 \citep[e.g.][]{prato03}.  Converting the 2MASS $K_s$ magnitudes to the Bessell-Brett homogenized system \citep{carpenter01} and averaging the two $K$-band magnitudes from Table~\ref{tab.compmag}, we obtain dereddened $K-L$ = $0.86 \pm 0.14$ mag for V807 Tau A and $0.31 \pm 0.27$ mag and $0.05 \pm 0.35$ mag for Ba and Bb, respectively.  As with the SED fits, these values show strong evidence for a disk in V807 Tau A and the absence of a disk around Ba and Bb.

\subsection{Comparison of Radio and Infrared Positions}
\label{sect.ir_radio}

We detected a compact radio source in three of the six epochs of VLBA observations.  Using the optical coordinates and proper motion of V807 Tau in \citet{ducourant05}, we computed the location of the optical source during the times of the VLBA epochs.  Assuming that the optical coordinates are dominated by the light from V807 Tau A and using the relative positions of V807 Tau A, Ba, and Bb from the adaptive optics images, we find that the VLBA coordinates are consistent with the location of the close pair to within $\sim$ 50 mas.  The standard error on the optical coordinates is $\sim$ 39 mas \citep{ducourant05}.  

To determine whether the radio emission is coming from V807 Tau Ba or Bb, we used our astrometric orbital fit (Fig.~\ref{fig.astrom}) to compute the predicted locations of Ba and Bb during the times of the VLBA observations.  We then applied the appropriate parallactic shift caused by a parallax of 7.14 mas (using the 140 pc distance from Kenyon et al.\ 1994) and added in the proper motion to compare the infrared positions with the motion observed in the VLBA reference frame.  During the two epochs with a detection of only a single source, we find that the radio position is consistent with the infrared location of V807 Tau Ba on 2007 Mar 10 and 2008 Aug 28.  On 2008 Mar 19, the faint radio source agrees with the infrared location of V807 Tau Ba while the brighter radio source matches the location of V807 Tau Bb.  The relative position of the two radio sources during this epoch ($\rho = 37.75$ mas, PA = 75.04$^\circ$) agrees quite nicely with the predicted position based on our orbit fit ($\rho = 37.80$ mas, PA = 75.00$^\circ$).

%We find that the radio positions are consistent with the infrared location of V807 Tau Ba on 2007 Mar 10 and 2008 Aug 28 and that it matches the location of V807 Tau Bb on 2008 Mar 19.  

To minimize the difference between the infrared and radio positions, we adjusted the parallax, proper motion, and the relative shift between the two frames.  Following this approach, we determined a best fitting parallax of $\pi = 7.3 \pm 1.2$ and proper motion of $\mu_\alpha\cos{\delta} = 7.6 \pm 1.8$ mas\,yr$^{-1}$, $\mu_\delta = -13.0 \pm 1.4$ mas\,yr$^{-1}$.  We show a comparison of the radio and infrared positions using these proper motions in Figure~\ref{fig.vlba_motion}.  Our measurements are consistent with the range of proper motions measured for stars in the Taurus star forming region \citep{ducourant05,bertout06}.  They are marginally consistent with the proper motion measured individually for V807 Tau by \citet{ducourant05} ($\mu_\alpha\cos{\delta} = 10$ mas\,yr$^{-1}$, $\mu_\delta = -20$ mas\,yr$^{-1}$).  The optical measurement could be affected by an offset in the photocenter caused by the orbital motion of the three components in the system which might account for the discrepancy in $\mu_\delta$.  We note that V807 Tau lies within the same region of the Taurus cloud complex as HP Tau/G2, a target that has a reliable parallax and proper motion measured using the VLBA \citep[see spatial distribution of VLBA sources in Fig.\ 5 of][]{torres09}.  The proper motion we measure for V807 Tau is similar to that measured for HP Tau/G2 ($\mu_\alpha\cos{\delta} = 13.85 \pm 0.03$ mas\,yr$^{-1}$, $\mu_\delta = -15.4 \pm 0.02$ mas\,yr$^{-1}$, $\pi = 6.20 \pm 0.03$ mas), as would be expected from two sources that might be located physically near to each other.

%Best pm_RA:       7.6000001   +        1.8000000   -        1.8000000
%Best pm_DEC:       -13.00000   +        1.4000000   -        1.4000000
%Best par:       7.3400000   +        1.2000000   -        1.1000000

% d = 161.2 +/- 0.9 pc

%Following this approach, we determined a best fitting proper motion of $\mu_\alpha\cos{\delta} = 8.0 \pm 2.0$ mas\,yr$^{-1}$, $\mu_\delta = -12.7 \pm 1.8$ mas\,yr$^{-1}$.

Ultimately, our goal is to combine these multi-epoch VLBA observations of V807 Tau to derive a distance with 2$-$4\% precision, similar to that obtained for other young stars in Taurus and Ophiuchus \citep{lestrade99,loinard05,loinard08,torres07,torres09,torres12}.  Our preliminary parallax has a much larger uncertainty, but is consistent with the range of distances measured for other members of the Taurus star forming region (130$-$160 pc).  It was difficult to obtain a more precise parallax for V807 Tau based on the current set of VLBA data because the radio emission from the two components is weak and variable and was not detected in half of the epochs.  We plan to continue these observations in the future to improve the precision of our results.

Based on the VLBA maps, the size of the radio emitting regions during each epoch is smaller than 2 mas ($\lesssim$ 60 R$_\odot$ or $\lesssim$ 40 R$_*$ assuming a stellar radius of 1.5 R$_\odot$).  The lack of disk signatures associated with V807 Tau Ba and Bb suggests that the variable and compact radio emission likely arises from magnetic activity and reconnection flares in the stellar magnetospheres of each component \citep[e.g.][]{feigelson99}.  The properties of the radio emission from V807 Tau Ba and Bb are consistent with those found around other weak-lined T Tauri stars which often show variability and the presence of radio flares \citep[e.g.][]{oneal90,feigelson94}.  The compact size and correspondingly high brightness temperatures of radio emission associated with weak-lined T Tauri stars generally indicates a non-thermal emission source \citep{phillips91,andre92}.  Using the major axis of the Gaussian models that we fit to the VLBA data as an upper limit to the size of the radio emitting regions around V807 Tau Ba and Bb, we compute a brightness temperature ranging from $5.3\times10^6$ to $1.3\times10^8$ \citep[using Eq.\ 1 in][]{skinner93}; this is consistent with the brightness temperatures expected from non-thermal emission \citep[$\gtrsim 10^{7}$;][]{andre92}.  Circular polarization is often undetected in radio emitting weak-lined T Tauri stars, but has been detected occasionally in a few of the stronger radio emitting weak-lined T Tauri stars, typically at a level of a few percent \citep[e.g.][]{white92b}.  The presence of circular polarization in the stronger sources indicates that the emission is likely gyrosynchrotron radiation.  The upper limit on the degree of polarization measured for V807 Tau ($<$ 8\%) is consistent with these expectations.  In another young triple system, the close binary V773 Tau Aa$-$Ab ($P \sim$ 51 days, $a \sim$ 0.3$-$0.5 AU) shows enhanced radio flares during periastron passage \citep{massi06,massi08,torres12}.  However, the larger separation between the components in V807 Tau Ba$-$Bb ($\sim$ 3.8$-$7.0 AU) makes it unlikely that similar inter-binary magnetic interactions would take place in the system.

%The discovery of circular polarization around stronger radio emitting weak-lined T Tauri stars indicates that the emission is likely gyrosynchrotron radiation \citep[e.g.][]{white92b}.  The low degree of polarization of only a few percent measured in some of the detected sources is consistent with the upper bound on V807 Tau ($<$ 8\%).  

%The lack of disk signatures within V807 Tau B suggests that the compact radio emission likely arises from unresolved flares inside the coronal zones of the components \citep[e.g.][]{phillips91}.  Interestingly, for the three PMS hierarchical triples that have been monitored with the VLBA or VLBI (V773 Tau, T Tau, and V807 Tau), the compact radio emission seems to be associated with the close binary pairs \citep{lestrade99,loinard08,torres12}.  Two other tight binaries (S1 and DoAr 21) also show radio emission detected with the VLBA \citep{loinard08}.  The association of radio emission with close binaries could indicate that magnetic field interactions between the components might be driving the emission \citep[e.g.][]{feigelson94}.  However, we note that this is not exclusively the case; there are stars that are not members of close binaries (Hubble 4, HD 283572, and HP Tau/G2) where radio emission has been observed and mapped in VLBA parallax programs \citep{torres07,torres09}\footnote{\citet{torres09} note that residuals in the astrometric motion of Hubble 4 could indicate the presence of an unseen companion.}.

\section{Discussion}

\subsection{Comparison with evolutionary tracks}
\label{sect.tracks}

We compare the locations of V807 Tau A, Ba, and Bb in the HR diagram to theoretical models of pre-main sequence stellar evolution in Figure~\ref{fig.hrd}.  We plot the luminosity and effective temperature of each component listed in Table~\ref{tab.sed} and compute expected masses and ages based on the evolutionary tracks computed by \citet[][BCAH]{baraffe98}, \citet[][SDF]{siess00}, and \citet[][Pisa models]{tognelli11}.  

The BCAH tracks predict a mass of $0.58 \pm 0.13$ M$_\odot$ at an age of $2.5 \pm 0.5$ Myr for V807 Tau Ba and Bb and a mass of $0.76 \pm 0.15$ M$_\odot$ at an age of $0.9 \pm 0.2$ Myr for V807 Tau A.  The SDF tracks predict a mass of $0.38 \pm 0.08$ M$_\odot$ at an age of $2.5 \pm 0.5$ Myr for V807 Tau Ba and Bb and a mass of $0.76 \pm 0.15$ M$_\odot$ at an age of $2.0 \pm 1.0$ Myr for V807 Tau A.    For the Ba and Bb components, the BCAH tracks predict a mass greater than the SDF tracks while the age estimates are about the same.  Compared with the dynamical mass measurements at a distance of 140 pc, the BCAH tracks are in good agreement with the mass of Ba ($M_{\rm Ba} =  0.564~(\frac{d}{\rm 140 pc})^3$ M$_\odot$), while the SDF tracks underestimate its value by 2\,$\sigma$.  For the Bb component ($M_{\rm Bb} =  0.476~(\frac{d}{\rm 140 pc})^3 $ M$_\odot$), the BCAH tracks overestimate the dynamical mass while the SDF tracks underestimate the value, both by $\sim$ 1\,$\sigma$.  The precision in determining the effective temperature based on our library of spectral templates limits our ability to discern finer differences in the predicted masses of V807 Tau Ba and Bb from the evolutionary tracks.  Furthermore, if we include the uncertainty in the distance to V807 Tau, then each of the predicted evolutionary masses becomes consistent with the dynamical estimates to within 1\,$\sigma$.  The uncertainty in the distance can be improved independently, for example, by measuring a geometric parallax using the VLBA.  The internal precision of the dynamical masses from the orbit fitting indicates that with an improved estimate of the distance, we can begin to make more meaningful comparisons between the evolutionary tracks.

For the Pisa models, we compare our results with tracks at solar metallicity ($Z$=0.02, $Y$=0.288, deuterium fraction 2$\times$10$^{-4}$) using a mixing length parameter of $\alpha = 1.2$ and 1.68.  For $\alpha = 1.2$, the Pisa tracks predict a mass of $0.46 \pm 0.10$ M$_\odot$ at an age of $2.0 \pm 1.0$ Myr for V807 Tau Ba and Bb and a mass of $0.88 \pm 0.18$ M$_\odot$ at an age of $2.0 \pm 1.0$ Myr for V807 Tau A.  For the larger mixing length of $\alpha = 1.68$, the Pisa tracks predict a mass of $0.40 \pm 0.09$ M$_\odot$ at an age of $2.5 \pm 0.5$ Myr for V807 Tau Ba and Bb and a mass of $0.68 \pm 0.18$ M$_\odot$ at an age of $1.8 \pm 0.5$ Myr for V807 Tau A.  Based on these results, it appears that the tracks at $\alpha = 1.2$ are more consistent with the dynamical masses of V807 Tau Ba and Bb at a distance of 140 pc.  As mentioned previously, including the current uncertainty in the distance to V807 Tau makes this distinction less significant.

For the evolutionary models that we considered, the ages for the components of V807 Tau are approximately 2 Myr.  This is in agreement with the mean age of 2.5 Myr determined from a larger census of the Taurus star forming region in \citet{white01}.  If all three components in V807 Tau formed at the same time, the agreement among the ages predicted for each component is more consistent for the SDF and Pisa tracks than it is for the BCAH models.  The age of the system depends on the actual distance to V807 Tau.  Because the distance provides the same scaling factor for all three components, the relative coevality between the components should remain consistent for a given set of tracks while the absolute age will shift.  For members of the Taurus star forming region with accurate VLBA parallaxes, the spread in distance ranges from 130 to 160 pc \citep{torres09}.  Our comparisons assume a distance of 140 pc.  If the distance proves to be closer to 160 pc (as the location of V807 Tau in Fig. 5 of Torres et al. 2009 might suggest), then the luminosities will be 0.12 dex greater and the ages would turn out to be closer to 1 Myr.

\subsection{Component Rotation}
\label{sect.rot}

Assuming that the rotational axes of V807 Tau A,
Ba, Bb, and the orbital axis of the Ba$-$Bb binary are aligned,  the
projected rotational velocities we measured imply that Ba
and Bb ($v\sin{i}$ = 60$-$70 km\,s$^{-1}$) are rotating much faster than A ($v\sin{i}$ = 15 km\,s$^{-1}$).  As discussed in Section~\ref{sect.sed}, V807 Tau A is identified as a classical T Tauri star with an infrared excess from a circumstellar disk, while V807 Tau Ba and Bb are classified as weak-lined T Tauri stars without any evidence for a disk.
The most efficient method for young stars to shed angular momentum is
through disk locking \citep[e.g.][]{stassun03,dahm12}. Therefore, the
presence of a disk around V807 Tau A and none within the V807 Tau B pair 
may explain the large difference in the rotational velocities
between the components.  The difference in rotational velocities in the V807 Tau system is also consistent with the study by \citet{meibom07} which shows that young, close binaries typically rotate faster than single stars and wider binaries, as a consequence of the close companion shortening the lifetime of the circumstellar disk and, in turn, the amount of time available for disk-braking.

%V807 Tau A is classified as a classical T Tauri star showing signs of an accretion disk, while the unresolved spectrum of V807 Tau Ba and Bb is classified as a weak-lined T Tauri star \citep{white01,hartigan03}. 

%"The observed effect of on average faster rotation in close binaries is consistent with a model scenario involving truncation of the CS disk lifetime by a close companion and consequently a shortened phase of magnetic disk-braking of the stellar rotation during the early PMS phase."

The break-up velocity of a 0.5~$M_\odot$ star is $\sim 203$ km\,s$^{-1}$ 
using a radius of $\sim 1.5$ $R_\odot$ at 2 Myr \citep{siess00}.  If the rotational axes of Ba and Bb are inclined by about 
the same amount as the orbital inclination of the Ba$-$Bb binary, $i\sim
152^\circ$, then they are actually rotating about twice as fast, $v \sim 140$ km\,s$^{-1}$, or about two thirds of breakup.  
The rotational velocity measured for V807 Tau Ba and Bb is higher than the average value of $v\sin{i} \sim$ 11 km\,s$^{-1}$ measured for M-stars in the Taurus star forming region \citep{nguyen09}, however there are individual low-mass young stars which have projected rotational velocities similar to V807 Tau Ba and Bb.  To remove the dependence on the inclination from this comparison, we can convert the estimated rotational velocity of the components in V807 Tau Ba and Bb to a rotation period of $\sim$ 0.5 days (using $R_* \sim 1.5~R_\odot$) and compare to the rotation periods measured for stars in young clusters.  As shown in \citet{irwin09}, although it is rare for low-mass young stars to be rotating so rapidly, there are other $\sim$ 0.5~$M_\odot$ stars present in nearby young clusters that have similarly short rotation periods at an age of about 2$-$5 Myr.

\section{Summary}

1. We measured the orbital motion within the PMS triple, V807 Tau, using adaptive optics imaging at the Keck Observatory.  The relative orbit of V807 Tau Ba$-$Bb combined with the motion of Ba and Bb around their center of mass, measured relative to V807 Tau A, yields dynamical masses of $M_{\rm Ba} =  0.564 \pm 0.018~(\frac{d}{\rm 140 pc})^3$ M$_\odot$ and  $M_{\rm Bb} =  0.476 \pm 0.017~(\frac{d}{\rm 140 pc})^3$ M$_\odot$.  The uncertainty in the masses represent the errors computed from the orbit fit; the systematic uncertainty in the distance is reflected by the additional scaling factor.

%The errors report only the uncertainty internal to the orbit fit; whereas the uncertainty in the distance is presented as a scaling factor.
%The uncertainties on the masses represent the errors computed from the orbit fit and do not account for the uncertainty in the distance to V807 Tau.

2. Based on spatially resolved high resolution spectra, we determined the spectral types of K7 for V807 Tau A and M2 for V807 Tau Ba and Bb.  The broadening of the spectral lines indicates that V807 Tau A is rotating with a projected velocity of 15 km~s$^{-1}$.  The spectra of V807 Tau Ba and Bb are rotationally broadened by 65$-$70 km~s$^{-1}$.  If the rotational axes of the three stars are aligned, this indicates that the components in the close pair are rotating much more quickly than V807 Tau A.

3. The spectral energy distribution of V807 Tau A shows an excess above the stellar model in the $K$ and $L$-bands.  This is consistent with its classification as a classical T Tauri star with an accretion disk.  The spectral energy distributions for V807 Tau Ba and Bb show that the infrared magnitudes are well matched by the spectral templates, consistent with a lack of circumstellar material around these weak-lined T Tauri stars.

4. Uncertainties in the distance to the system and the effective temperatures of the components are the main limitations to a more detailed comparison of our dynamical mass measurements with predictions based on evolutionary tracks for pre-main sequence stars.  At a distance of 140~pc, the tracks indicate an age of $\sim$ 2 Myr for the V807 Tau triple. 

%Our dynamical mass measurements for V807 Tau Ba and Bb are marginally consistent with predictions based on evolutionary tracks for pre-main sequence stars computed by \citet{baraffe98}, \citet{siess00}, and \citet{tognelli11}.  Uncertainties in the effective temperatures and the distance to the system are the main limitations to a more thorough comparison between the different models.  Overall, the tracks indicate an age of $\sim$ 2 Myr for the V807 Tau triple.

5. We detected compact radio emission from V807 Tau during three out of six epochs using the VLBA.  The emission is variable, with the integrated intensity ranging from 0.52 to 8.60 mJy.  Comparing the position of the radio source with the motion observed in our infrared images, we determined that the emission originated from V807 Tau Ba in the first and third detected epochs, while both V807 Tau Ba and Bb were detected in the second epoch.  Continued monitoring with the VLBA in the future will provide a precise parallax measurement for V807 Tau.  This will help place the dynamical masses on an absolute scale.

\acknowledgments

We thank Vivek Dhawan, Amy Mioduszewski, and Walter Brisken for valuable advice and help in planning the VLBA observations.  We also appreciate the operational and instrument support provided at the Keck Observatory.  We thank the anonymous referee who provided valuable comments that helped improve the paper.  GHS acknowledges support from NASA Keck PI Data Awards administered by the NASA Exoplanet Science Institute (JPL contracts 1327033 and 1441975) and from NASA grant NNX07A176G.  The work of MS was supported in part by NSF grant AST-09-08406; NSF grant 10-09136 supported the contributions of LP.  Data were obtained at the W. M. Keck Observatory through time allocated by NASA through a partnership with Caltech and the University of California and by NOAO through the Telescope System Instrumentation Program (TSIP; 08B-0336, 09B-0040, 10B-0255). TSIP is funded by the NSF.  The Keck Observatory was made possible by the generous financial support of the W. M. Keck Foundation.  We recognize and acknowledge the significant cultural role that the summit of Mauna Kea plays within the indigenous Hawaiian community and are grateful for the opportunity to conduct these observations from the mountain.  The National Radio Astronomy Observatory is a facility of the National Science Foundation operated under cooperative agreement by Associated Universities, Inc.  This research has made use of data products from the Two Micron All Sky Survey and catalogs accessible through the VizieR catalogue access tool, CDS, Strasbourg, France.  The NIRSPEC data on 2003 Dec 13 (PI: A. Ghez) was retrieved through the Keck Observatory Archive (KOA), which is operated by the W. M. Keck Observatory and the NASA Exoplanet Science Institute (NExScI), under contract with the National Aeronautics and Space Administration.  This research also made use of data from the MOJAVE database that is maintained by the MOJAVE team.

{\it Facilities:} \facility{Keck:II (NIRC2, NIRSPEC)}, \facility{VLBA}

\clearpage

\clearpage

\begin{deluxetable}{llcccc}
\tablewidth{0pt}
\tablecaption{Log of Recent NIRC2 Adaptive Optics Observations\tablenotemark{a}}
\tablehead{\colhead{Date} & \colhead{Filter} & \colhead{AO Rate} & \colhead{T$_{int}$} & \colhead{No.} & \colhead{No.} \\ \colhead{} & \colhead{} & \colhead{(Hz)} & \colhead{(s)} & \colhead{Co-add} & \colhead{Im.}}
\startdata
2006 Dec 18  & Jcont & 492 & 1.00 & 10 & 10 \\ 
             & Hcont & 492 & 0.60 & 10 & 10 \\
             & Kcont & 492 & 0.50 & 10 & 10 \\
2008 Jan 17  & Hcont & 438 & 0.40 & 10 & 20 \\
             & Kcont & 438 & 0.40 & 10 & 20 \\
             & L$'$    & 438 & 0.05 & 10 & 10 \\ 
2008 Dec 17  & Hcont & 438 & 0.30 & 10 & 10 \\ 
             & Kcont & 438 & 0.30 & 10 & 10 \\
2009 Oct 25  & Hcont & 438 & 0.18 & 10 & 12 \\
             & Kcont & 438 & 0.18 & 10 & 12 \\
2011 Jan 24  & Hcont & 438 & 0.50 & 10 & 12 \\
             & Kcont & 438 & 0.50 & 10 & 12 \\
2011 Oct 12  & Hcont & 438 & 0.18 & 10 & 24 \\
             & Kcont & 438 & 0.18 & 10 &  6 \\
             & Kcont & 438 & 0.30 & 10 & 12
\enddata
\tablenotetext{a}{See \citet{schaefer06} for a log of observations obtained prior to 2006.}
\label{tab.ao-obs}
\end{deluxetable}

\clearpage

\begin{deluxetable}{llllll} 
\tabletypesize{\small}
\tablewidth{0pt}
\tablecaption{Relative positions of V807 Tau from AO imaging} 
\tablehead{
\colhead{Year} & \colhead{$\rho$(mas)} & \colhead{P.A.($\degr$)} & \colhead{Ref}}
\startdata 
\multicolumn{4}{c}{V807 Tau A$-$Ba} \\
 \noalign{\vskip .8ex}%
 \hline
 \noalign{\vskip .8ex}%
1999.7558  &  282.91  $\pm$  2.10  &  324.31  $\pm$  0.43  &  1  \\   %F 1999-10-03
2000.6565  &  286.86  $\pm$  1.80  &  323.56  $\pm$  0.36  &  1  \\   %F 2000-08-27
2000.6786  &  284.88  $\pm$  2.70  &  323.60  $\pm$  0.54  &  1  \\   %F 2000-09-04
2001.0006  &  287.00  $\pm$  1.80  &  322.83  $\pm$  0.36  &  1  \\   %F 2000-12-31
2001.1721  &  286.23  $\pm$  0.91  &  322.12  $\pm$  0.18  &  1  \\   %F 2001-03-04
2002.1829  &  285.57  $\pm$  0.61  &  320.34  $\pm$  0.12  &  1  \\   %A 2002-03-08
2002.8293  &  282.71  $\pm$  0.34  &  318.90  $\pm$  0.07  &  1  \\   %A 2002-10-30
2002.9750  &  281.40  $\pm$  1.50  &  319.25  $\pm$  0.31  &  1  \\   %F 2002-12-22
2002.9852  &  281.82  $\pm$  1.10  &  319.60  $\pm$  0.22  &  1  \\   %F 2002-12-26
2003.8539  &  271.57  $\pm$  1.01  &  317.33  $\pm$  0.21  &  1  \\   %F 2003-11-08
2003.9766  &  273.21  $\pm$  1.00  &  316.96  $\pm$  0.21  &  1  \\   %F 2003-12-23
2005.1644  &  268.90  $\pm$  1.00  &  312.69  $\pm$  0.22  &  1  \\   %A 2005-03-01
2005.5989  &  260.51  $\pm$  0.70  &  310.92  $\pm$  0.15  &  1  \\   %F 2005-08-06
2005.9317  &  259.77  $\pm$  0.31  &  309.30  $\pm$  0.07  &  1  \\   %A 2005-12-06
2006.9633  &  248.55  $\pm$  0.47  &  305.51  $\pm$  0.11  &  2  \\   %A 2006-12-18
2008.0447  &  234.61  $\pm$  0.28  &  301.43  $\pm$  0.07  &  2  \\   %A 2008-01-17
2008.9621  &  221.22  $\pm$  0.25  &  298.08  $\pm$  0.07  &  2  \\   %A 2008-12-17
2009.8167  &  208.08  $\pm$  0.45  &  295.23  $\pm$  0.12  &  2  \\   %A 2009-10-25
2011.0644  &  190.50  $\pm$  0.39  &  292.97  $\pm$  0.12  &  2  \\   %A 2011-01-24
2011.7797  &  183.43  $\pm$  0.77  &  292.59  $\pm$  0.24  &  2  \\   %A 2011-10-12
\cutinhead{V807 Tau A$-$Bb}                                                          
1999.7558  &  305.00  $\pm$  4.40  &  321.29  $\pm$  0.83  &  1  \\  %F 1999-10-03
2000.6565  &  284.59  $\pm$  4.10  &  316.81  $\pm$  0.83  &  1  \\  %F 2000-08-27
2000.6786  &  285.22  $\pm$  4.60  &  317.56  $\pm$  0.92  &  1  \\  %F 2000-09-04
2001.0006  &  282.86  $\pm$  4.20  &  315.60  $\pm$  0.85  &  1  \\  %F 2000-12-31
2001.1721  &  279.36  $\pm$  3.60  &  314.66  $\pm$  0.74  &  1  \\  %F 2001-03-04
2002.1829  &  259.74  $\pm$  0.93  &  313.81  $\pm$  0.20  &  1  \\  %A 2002-03-08
2002.8293  &  249.13  $\pm$  0.51  &  312.57  $\pm$  0.12  &  1  \\  %A 2002-10-30
2002.9750  &  247.05  $\pm$  2.20  &  312.80  $\pm$  0.51  &  1  \\  %F 2002-12-22
2002.9852  &  248.86  $\pm$  2.40  &  313.14  $\pm$  0.55  &  1  \\  %F 2002-12-26
2003.8539  &  227.53  $\pm$  2.40  &  312.98  $\pm$  0.60  &  1  \\  %F 2003-11-08
2003.9766  &  228.83  $\pm$  2.40  &  313.58  $\pm$  0.60  &  1  \\  %F 2003-12-23
2005.1644  &  220.81  $\pm$  1.26  &  311.63  $\pm$  0.33  &  1  \\  %A 2005-03-01
2005.5989  &  214.79  $\pm$  1.30  &  311.32  $\pm$  0.35  &  1  \\  %F 2005-08-06
2005.9317  &  215.27  $\pm$  0.46  &  310.55  $\pm$  0.12  &  1  \\  %A 2005-12-06
2006.9633  &  209.75  $\pm$  0.65  &  309.69  $\pm$  0.18  &  2  \\  %A 2006-12-18
2008.0447  &  207.72  $\pm$  0.37  &  308.48  $\pm$  0.10  &  2  \\  %A 2008-01-17
2008.9621  &  208.03  $\pm$  0.31  &  306.87  $\pm$  0.08  &  2  \\  %A 2008-12-17
2009.8167  &  209.40  $\pm$  0.72  &  304.15  $\pm$  0.20  &  2  \\  %A 2009-10-25
2011.0644  &  210.75  $\pm$  0.81  &  298.07  $\pm$  0.22  &  2  \\  %A 2011-01-24
2011.7797  &  208.02  $\pm$  1.13  &  293.02  $\pm$  0.31  &  2  \\  %A 2011-10-12
\cutinhead{V807 Tau Ba$-$Bb}                                                              
1999.7558  &   26.97  $\pm$  4.88  &  287.78  $\pm$ 10.37  &  1  \\  %FGS    F 1999-10-03
2000.6565  &   33.72  $\pm$  4.48  &  226.32  $\pm$  7.61  &  1  \\  %FGS    F 2000-08-27
2000.6786  &   30.01  $\pm$  5.33  &  231.24  $\pm$ 10.18  &  1  \\  %FGS    F 2000-09-04
2001.0006  &   36.16  $\pm$  4.57  &  222.65  $\pm$  7.24  &  1  \\  %FGS    F 2000-12-31
2001.1721  &   37.42  $\pm$  3.71  &  217.83  $\pm$  5.68  &  1  \\  %FGS    F 2001-03-04
2002.1829  &   40.38  $\pm$  1.36  &  187.39  $\pm$  1.93  &  1  \\  %NIRC2  A 2002-03-08
2002.8293  &   44.55  $\pm$  0.34  &  176.91  $\pm$  0.44  &  1  \\  %NIRC2  A 2002-10-30
2002.9750  &   45.40  $\pm$  2.66  &  176.98  $\pm$  3.36  &  1  \\  %FGS    F 2002-12-22
2002.9852  &   44.46  $\pm$  2.64  &  178.63  $\pm$  3.40  &  1  \\  %FGS    F 2002-12-26
2003.8539  &   47.90  $\pm$  2.60  &  158.42  $\pm$  3.11  &  1  \\  %FGS    F 2003-11-08
2003.9766  &   46.78  $\pm$  2.60  &  153.75  $\pm$  3.18  &  1  \\  %FGS    F 2003-12-23
2004.9814  &   47.80  $\pm$  1.72  &  138.71  $\pm$  2.06  &  1  \\  %NIRC2  A 2004-12-24 - El12N saturated
2005.1644  &   48.28  $\pm$  1.46  &  137.54  $\pm$  1.73  &  1  \\  %NIRI   A 2005-03-01
2005.5989  &   45.76  $\pm$  1.48  &  129.03  $\pm$  1.85  &  1  \\  %FGS    F 2005-08-06
2005.9317  &   44.80  $\pm$  0.47  &  123.28  $\pm$  0.61  &  1  \\  %NIRC2  A 2005-12-06
2006.9633  &   42.22  $\pm$  0.49  &  104.29  $\pm$  0.67  &  2  \\  %NIRC2  A 2006-12-18
2008.0447  &   38.22  $\pm$  0.36  &   79.57  $\pm$  0.54  &  2  \\  %NIRC2  A 2008-01-17
2008.9621  &   35.42  $\pm$  0.41  &   54.28  $\pm$  0.67  &  2  \\  %NIRC2  A 2008-12-17
2009.8167  &   32.47  $\pm$  0.41  &   27.38  $\pm$  0.73  &  2  \\  %NIRC2  A 2009-10-25
2011.0644  &   26.99  $\pm$  0.51  &  336.96  $\pm$  1.08  &  2  \\  %NIRC2  A 2011-01-24
2011.7797  &   24.64  $\pm$  0.52  &  296.16  $\pm$  1.22  &  2      %NIRC2  A 2011-10-12
\enddata 
\tablecomments{References: (1) Schaefer et al. 2006; (2) This paper.}
\label{tab.sepPA}
\end{deluxetable} 

\clearpage

\begin{deluxetable}{lcccc} 
\tablewidth{0pt}
\tablecaption{Flux Ratios Measured for V807 Tau} 
\tablehead{\colhead{Date} & \colhead{Filter} & \colhead{Ba/A} & \colhead{Bb/A} & \colhead{Bb/Ba}}
\startdata 
2002/03/08 &     H &  0.388 $\pm$ 0.017  &  0.279 $\pm$ 0.008  &  0.720 $\pm$ 0.043 \\
2002/03/08 &  K$'$ &  0.399 $\pm$ 0.011  &  0.276 $\pm$ 0.008  &  0.692 $\pm$ 0.026 \\
2002/10/30 &     H &  0.381 $\pm$ 0.016  &  0.266 $\pm$ 0.010  &  0.698 $\pm$ 0.010 \\
2002/10/30 &  K$'$ &  0.322 $\pm$ 0.010  &  0.233 $\pm$ 0.008  &  0.723 $\pm$ 0.008 \\
2004/12/24 &     H &  0.710 $\pm$ 0.048  &  0.556 $\pm$ 0.031  &  0.784 $\pm$ 0.045 \\
2004/12/24 &  K$'$ &  0.606 $\pm$ 0.053  &  0.452 $\pm$ 0.024  &  0.748 $\pm$ 0.043 \\
2005/03/01 & Hcont &  0.393 $\pm$ 0.009  &  0.271 $\pm$ 0.008  &  0.689 $\pm$ 0.032 \\
2005/03/01 & Kcont &  0.308 $\pm$ 0.012  &  0.224 $\pm$ 0.013  &  0.731 $\pm$ 0.074 \\
2005/12/06 &     H &  0.388 $\pm$ 0.003  &  0.284 $\pm$ 0.003  &  0.731 $\pm$ 0.008 \\
2005/12/06 &  K$'$ &  0.326 $\pm$ 0.008  &  0.240 $\pm$ 0.007  &  0.736 $\pm$ 0.004 \\
2006/12/18 & Jcont &  0.367 $\pm$ 0.010  &  0.277 $\pm$ 0.009  &  0.757 $\pm$ 0.037 \\
2006/12/18 & Hcont &  0.363 $\pm$ 0.007  &  0.256 $\pm$ 0.007  &  0.706 $\pm$ 0.027 \\
2006/12/18 & Kcont &  0.295 $\pm$ 0.005  &  0.223 $\pm$ 0.003  &  0.758 $\pm$ 0.019 \\
2008/01/17 & Hcont &  0.371 $\pm$ 0.006  &  0.252 $\pm$ 0.004  &  0.678 $\pm$ 0.007 \\
2008/01/17 & Kcont &  0.291 $\pm$ 0.003  &  0.212 $\pm$ 0.003  &  0.729 $\pm$ 0.008 \\
2008/01/17 &  L$'$ &  0.194 $\pm$ 0.027  &  0.109 $\pm$ 0.024  &  0.593 $\pm$ 0.230 \\
2008/12/17 & Hcont &  0.360 $\pm$ 0.005  &  0.243 $\pm$ 0.004  &  0.675 $\pm$ 0.009 \\
2008/12/17 & Kcont &  0.282 $\pm$ 0.002  &  0.199 $\pm$ 0.002  &  0.707 $\pm$ 0.012 \\
2009/10/25 & Hcont &  0.396 $\pm$ 0.004  &  0.282 $\pm$ 0.003  &  0.711 $\pm$ 0.009 \\
2009/10/25 & Kcont &  0.307 $\pm$ 0.009  &  0.230 $\pm$ 0.008  &  0.749 $\pm$ 0.048 \\
2011/01/24 & Hcont &  0.401 $\pm$ 0.006  &  0.265 $\pm$ 0.006  &  0.660 $\pm$ 0.023 \\
2011/01/24 & Kcont &  0.342 $\pm$ 0.007  &  0.219 $\pm$ 0.008  &  0.642 $\pm$ 0.037 \\
2011/10/12 & Hcont &  0.387 $\pm$ 0.004  &  0.273 $\pm$ 0.004  &  0.705 $\pm$ 0.011 \\
2011/10/12 & Kcont &  0.312 $\pm$ 0.020  &  0.204 $\pm$ 0.022  &  0.660 $\pm$ 0.111 \\
\hline
Weighted Mean & V  &  0.1336 $\pm$ 0.0035 & 0.0635 $\pm$ 0.0026 & 0.4688 $\pm$ 0.0221 \\
Weighted Mean & J  &  0.3666 $\pm$ 0.0105 & 0.2773 $\pm$ 0.0092 & 0.7572 $\pm$ 0.0373 \\
Weighted Mean & H  &  0.3849 $\pm$ 0.0016 & 0.2713 $\pm$ 0.0014 & 0.6995 $\pm$ 0.0034 \\
Weighted Mean & K  &  0.2956 $\pm$ 0.0016 & 0.2146 $\pm$ 0.0013 & 0.7313 $\pm$ 0.0028 \\
Weigthed Mean & L  &  0.194  $\pm$ 0.027  &  0.109 $\pm$ 0.024  &  0.593 $\pm$ 0.230 
\enddata
\label{tab.aoflux} 
\end{deluxetable}

\begin{deluxetable}{lcccccc} 
\tabletypesize{\small}
\tablewidth{0pt}
\tablecaption{Magnitudes of the Components in V807 Tau} 
\tablehead{\colhead{Component} & \colhead{V} & \colhead{J} & \colhead{H} & \colhead{K} & \colhead{L} & \colhead{ref}}
\startdata 
%             V                     J                    H                     K                    L
V807 Tau    & 11.185 $\pm$ 0.114  & 8.146 $\pm$ 0.023  & 7.357 $\pm$ 0.026   & 6.960 $\pm$ 0.016  & \ldots            &  1,2 \\
V807 Tau A  & 11.56  $\pm$ 0.02   & \ldots             & \ldots              & 7.36  $\pm$ 0.05   & 6.54 $\pm$ 0.14   &  3   \\
V807 Tau B  & 13.35  $\pm$ 0.04   & \ldots             & \ldots              & 8.21  $\pm$ 0.10   & 7.94 $\pm$ 0.14   &  3   \\
\hline                                                                                                 
V807 Tau A  & 11.380 $\pm$ 0.114  & 8.686 $\pm$ 0.025  & 7.905 $\pm$ 0.026   & 7.408 $\pm$ 0.016  & \ldots            &  4 \\
V807 Tau A  & 11.400 $\pm$ 0.390  & \ldots             & \ldots              & \ldots             & \ldots            &  5 \\                       
V807 Tau Ba & 13.566 $\pm$ 0.117  &  9.775 $\pm$ 0.034 & 8.941 $\pm$ 0.026   & 8.731 $\pm$ 0.017  & \ldots            &  4 \\
V807 Tau Ba & 13.767 $\pm$ 0.043  &  \ldots            & \ldots              & 8.806 $\pm$ 0.100  & 8.445 $\pm$ 0.220 &  6 \\
V807 Tau Ba & 13.572 $\pm$ 0.390  &  \ldots            & \ldots              & \ldots             & \ldots            &  5 \\
V807 Tau Bb & 14.373 $\pm$ 0.122  & 10.078 $\pm$ 0.039 &  9.321 $\pm$ 0.026  & 9.078 $\pm$ 0.017  & \ldots            &  4 \\
V807 Tau Bb & 14.590 $\pm$ 0.054  &  \ldots            & \ldots              & 9.146 $\pm$ 0.100  & 9.013 $\pm$ 0.334 &  6 \\
V807 Tau Bb & 14.368 $\pm$ 0.392  &  \ldots            &  \ldots             & \ldots             & \ldots            &  5 
\enddata
\tablecomments{References: (1) Kharchenko et al. 2009 ($V$-band); (2) Cutri et al. 2003 (2MASS $J$, $H$, $K_s$); (3) White \& Ghez 2001 ($V$, $K$); (4) Flux ratio (A:Ba:Bb) combined with Johnson V All-sky compiled Catalog (Kharchenko 2009) and 2MASS J, H, K; (5) Calibrated FGS; (6) Flux ratio (Ba:Bb) combined with White and Ghez (2001).} 
\label{tab.compmag} 
\end{deluxetable} 

\clearpage

\begin{deluxetable}{lccl}
\tablewidth{0pt}
\tablecaption{Log of NIRSPEC/NIRSPAO Observations}
\tablehead{\colhead{UT Date} & \colhead{Phase} & \colhead{Sep (mas)} & \colhead{Spectra}}
\startdata
2001 Jan 8  & 0.15  & 33 & A, Ba, Bb\\
2002 Jul 16 & 0.27  & 43 & blended, non-AO \\
2002 Dec 14 & 0.31  & 45 & blended, non-AO \\
2003 Dec 13 & 0.39  & 47 & A, Ba, Bb\\
2006 Dec 14 & 0.63  & 42 & A, Ba+Bb\\
2009 Dec 6  & 0.88  & 32 & Ba+Bb \\
2010 Dec 12 & 0.96  & 28 & A, Ba+Bb
\enddata
\label{tab.nirspec_obs}
\end{deluxetable}

\begin{deluxetable}{lcccccccc}
\tabletypesize{\small}
\tablewidth{0pt}
\tablecaption{NIRSPEC/NIRSPAO Results}
\tablehead{
\colhead{}        & \colhead{}               & \multicolumn{3}{c}{V807 Tau A} & \multicolumn{2}{c}{V807 Tau Ba} & \multicolumn{2}{c}{V807 Tau Bb} \\ \hline
\colhead{UT Date} & \colhead{JD - 2,400,000} & \colhead{SpT} & \colhead{$v~\sin{i}$} & \colhead{$v_{\rm rad} $} & \colhead{SpT} & \colhead{$v~\sin{i}$} & \colhead{SpT} & \colhead{$v~\sin{i}$} \\
\colhead{}        & \colhead{}               & \colhead{}  & \colhead{(km s$^{-1}$)} & \colhead{(km s$^{-1}$)} & \colhead{} & \colhead{(km s$^{-1}$)} & \colhead{} & \colhead{km s$^{-1}$} }
\startdata
2001 Jan 8  & 51917.8472 & K7 & 15 & $16.22 \pm 1.0$    &  M2 & $70\pm10$  & M2    & $70 \pm 10$ \\
2002 Jul 16 & 52472.1403 & K7 & 15 & $16.42 \pm 1.0$    &     &            &       &      \\
2002 Dec 14 & 52622.8764 & K7 & 15 & $16.35 \pm 1.0$    &     &            &       &      \\
2003 Dec 13 & 52986.8697 & K7 & 15 & $15.55 \pm 1.0$    &  M2 & $65\pm10$  & M2.5  & $60 \pm 10$ \\
2006 Dec 14 & 54083.9033 & K7 & 15 & $15.40 \pm 1.0$    &     &            &       &      \\
2010 Dec 12 & 55542.8255 & K7 & 15 & $14.80 \pm 1.0$    &     &            &       &      \\ 
\hline
Average     &            &    & 15 & $15.79 \pm 0.65$   &     & $68 \pm 7$ &       & $65 \pm 7$
\enddata
\label{tab.nirspec_vel}
\end{deluxetable}

\clearpage

\begin{deluxetable}{lccccccc}
\tabletypesize{\footnotesize}
%\tabletypesize{\small}
\tablewidth{0pt}
\tablecaption{VLBA Observations of V807 Tau}
\tablehead{\colhead{Date} & \colhead{Comp} & \colhead{RA} & \colhead{DEC} & \colhead{$f_\nu$} & \colhead{Major} & \colhead{Minor} & \colhead{PA} \\ \colhead{} & \colhead{} & \colhead{(hh:mm:ss)} & \colhead{(dd:mm:ss)} & \colhead{(mJy)} & \colhead{(mas)} & \colhead{(mas)} & \colhead{($^\circ$)}}
\startdata
2007 Mar 10 & Ba & 04 33 06.6247856 & 24 09 54.989774 & 0.52 $\pm$ 0.10 & 2.27 $\pm$ 0.29 & 1.18 $\pm$ 0.15 & 8.3 $\pm$ 7.4 \\        %BS171   & 
2008 Mar 19 & Ba & 04 33 06.6255328 & 24 09 54.953108 & 0.59 $\pm$ 0.13 & 2.11 $\pm$ 0.34 & 1.88 $\pm$ 0.30 & 48 $\pm$ 58  \\         %BS176A  & 
2008 Mar 19 & Bb & 04 33 06.6281980 & 24 09 54.962855 & 1.07 $\pm$ 0.09 & 1.90 $\pm$ 0.09 & 1.08 $\pm$ 0.05 & 178.5 $\pm$  3.3 \\     %BS176A  & 
2008 Jun 04 & \nodata &  \nodata    &    \nodata      & $<$ 0.22        &    \nodata      &     \nodata     & \nodata  \\             %BS176B  & 
2008 Aug 28 & Ba & 04 33 06.6270069 & 24 09 54.939863 & 8.60 $\pm$ 0.27 & 1.86 $\pm$ 0.03 & 1.00 $\pm$ 0.02 & 164.2 $\pm$ 1.1 \\      %BS176C  & 
2008 Dec 10 & \nodata &  \nodata    &    \nodata      & $<$ 0.27        &    \nodata      &     \nodata     & \nodata  \\             %BS176D  & 
2009 Mar 06 & \nodata &  \nodata    &    \nodata      & $<$ 0.26        &    \nodata      &     \nodata     &  \nodata                %BS176E  & 
\enddata
\label{tab.vlba}
\end{deluxetable}

\clearpage

\begin{deluxetable}{ll}
\tablewidth{0pt}
\tablecaption{Orbital Parameters and Masses for V807 Tau Ba$-$Bb}
\tablehead{\colhead{Parameter} & \colhead{Value}}
\startdata
$P$ (yrs)          &  12.312   $\pm$ 0.058  \\
$T$                &  1999.154 $\pm$ 0.070  \\
$e$                &  0.2922   $\pm$ 0.0033 \\
$a$ (mas)          &  38.59    $\pm$ 0.16   \\
$i$ ($^\circ$)      &  151.59   $\pm$ 0.89   \\
$\Omega$ ($^\circ$) &  1.8      $\pm$ 1.6    \\
$\omega$ ($^\circ$) &  50.6     $\pm$ 1.3    \\
$\chi^2$           &  15.49                 \\
$\chi^2_\nu$        &  0.443                 \\ 
$M_{\rm tot}~(\frac{d}{\rm 140 pc})^3 (M_\odot)$   &  1.040 $\pm$ 0.016  \\
$M_{\rm Bb}/M_{\rm Ba}$  &  0.843 $\pm$ 0.050  \\
$M_{\rm Ba}~(\frac{d}{\rm 140 pc})^3 (M_\odot)$   &  0.564 $\pm$ 0.018  \\
$M_{\rm Bb}~(\frac{d}{\rm 140 pc})^3 (M_\odot)$   &  0.476 $\pm$ 0.017
\label{tab.orbit}
\enddata
\tablecomments{The uncertainty in the masses represent the errors computed from the orbit fit; the systematic uncertainty in the distance is reflected by the additional scaling factor of $(\frac{d}{\rm 140~pc})^3$.  The uncertainty in the average distance to the Taurus star forming region \citep[$\pm$ 10 pc;][]{kenyon94} could shift the masses by as much as 0.24 M$_\odot$.}
\end{deluxetable}

\begin{deluxetable}{lcccccc}
\tabletypesize{\scriptsize}
\tablewidth{0pt}
\tablecaption{Spectral Energy Distributions}
\tablehead{\colhead{Star} & \colhead{SpT} & \colhead{$T_{\rm eff}$} & \colhead{$A_V$} &\colhead{$\theta$} & \colhead{$F_{\rm bol}$} & \colhead{$L_*$ ($\frac{d}{\rm 140 pc})^2$ } \\ 
\colhead{} & \colhead{} & \colhead{(K)} & \colhead{(mag)} &\colhead{(mas)} & \colhead{($10^{-9}$ erg/cm$^2$/s)} & \colhead{(L$_\odot$)} }
\startdata
V807 Tau A   & K7        & 4060  & 0.03 $\pm$ 0.42 & 0.14 $\pm$ 0.03 & 1.72 $\pm$ 0.32 &  1.05 $\pm$ 0.20 \\ % & SED, No K,L \\ 
V807 Tau Ba  & M2        & 3560  & 0.61 $\pm$ 0.50 & 0.12 $\pm$ 0.01 & 0.732 $\pm$ 0.094 & 0.447 $\pm$ 0.058 \\ % & SED, All \\
V807 Tau Bb  & M2$-$M2.5 & 3560  & 1.18 $\pm$ 0.47 & 0.11 $\pm$ 0.01 & 0.483 $\pm$ 0.063 & 0.356 $\pm$ 0.038 % & SED, All  
\label{tab.sed}
\enddata
\tablecomments{The uncertainty in the distance ($\pm$ 10 pc) contributes an additional systematic uncertainty of $\pm$ 0.16, 0.066, and 0.053 L$_\odot$ to the luminosities of A, Ba, and Bb, respectively.}
\end{deluxetable}

\clearpage

\clearpage

\begin{figure}
\begin{center}
   \scalebox{0.88}{\includegraphics{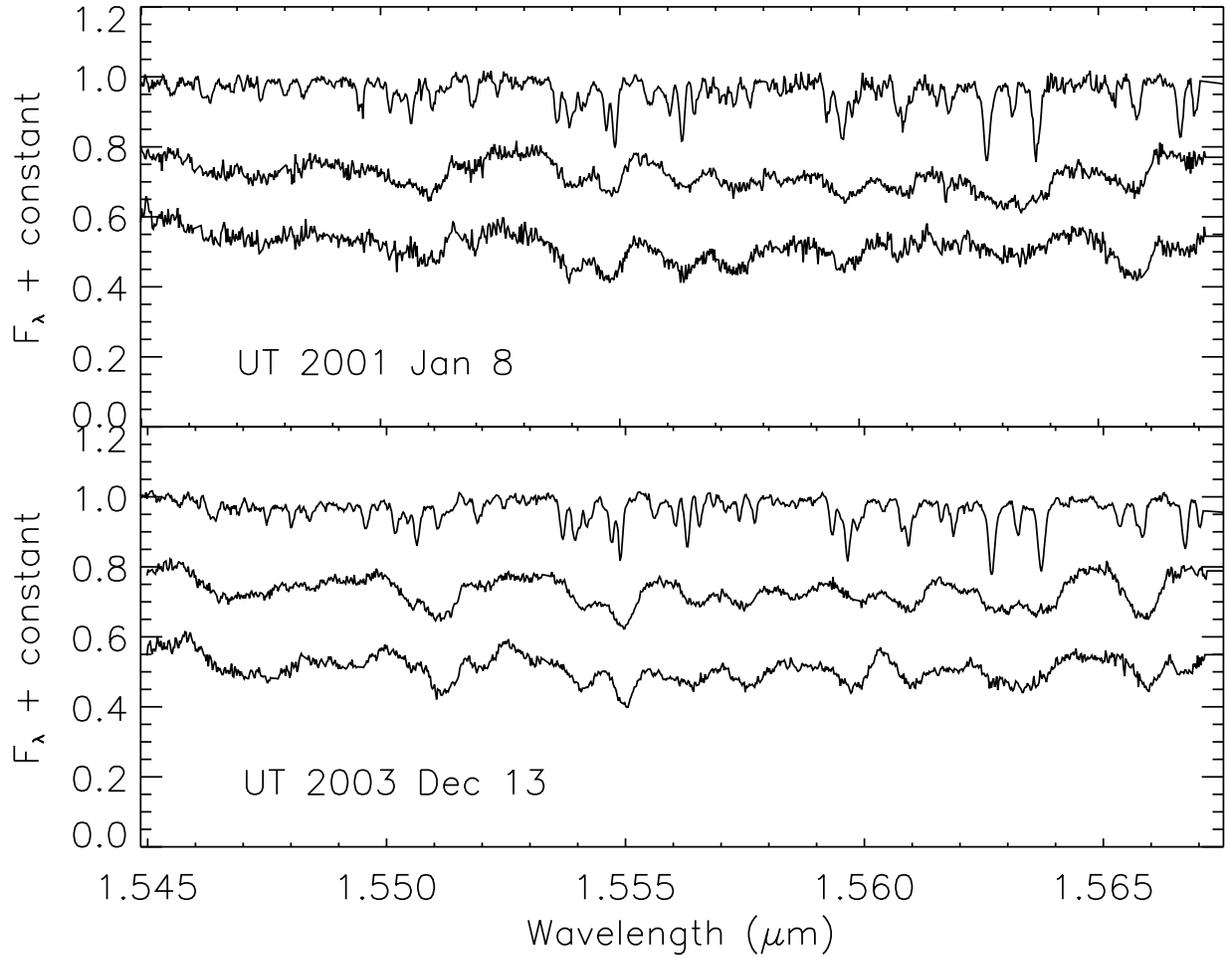}}
\end{center}
\caption{Angularly resolved spectra of V807 Tau A (top), Ba (middle), and Bb (bottom) at two epochs. The spectra are corrected for barycentric motion, normalized to 1.0, and offset by a constant. }%UT 2001 Jan 8  and UT 2003 Dec 12
\label{fig.spec}
\end{figure}

\clearpage

\begin{figure}
\begin{center}
   \scalebox{1.0}{\includegraphics{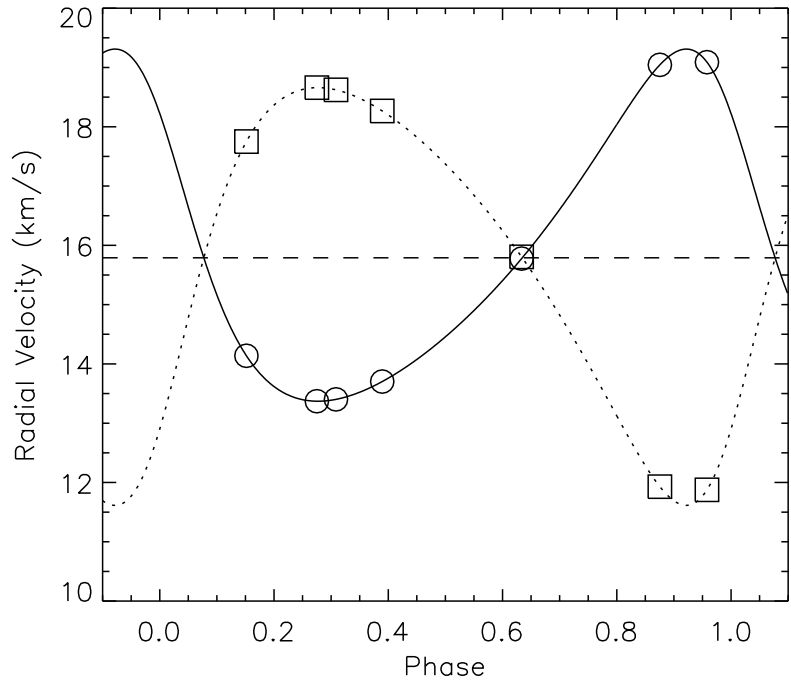}}
\end{center}
\caption{Predicted radial velocity curves for V807 Tau Ba (solid line) and Bb (dotted line) based on the orbital parameters and mass ratio from the visual and astrometric orbit.  The circles and squares show the expected radial velocities of Ba and Bb, respectively, during the times of the NIRSPEC observations.}
\label{fig.predRV}
\end{figure}

\clearpage

\begin{figure}
\begin{center}
  \scalebox{1.0}{\includegraphics{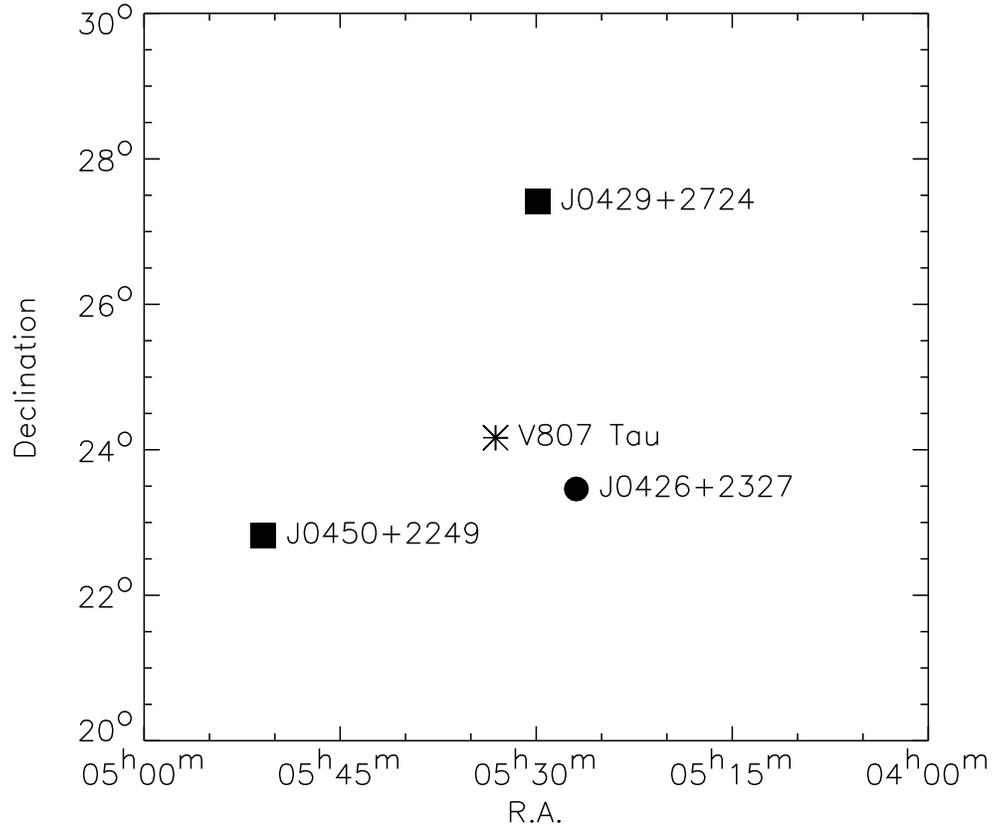}}
\end{center}
  \caption{VLBA source positions of V807 Tau relative to the primary phase reference calibrator (circle) and the secondary phase reference check sources (squares).}
\label{fig.vlbapos}
\end{figure}

\clearpage

\begin{figure}
\begin{center}
   \scalebox{0.52}{\includegraphics{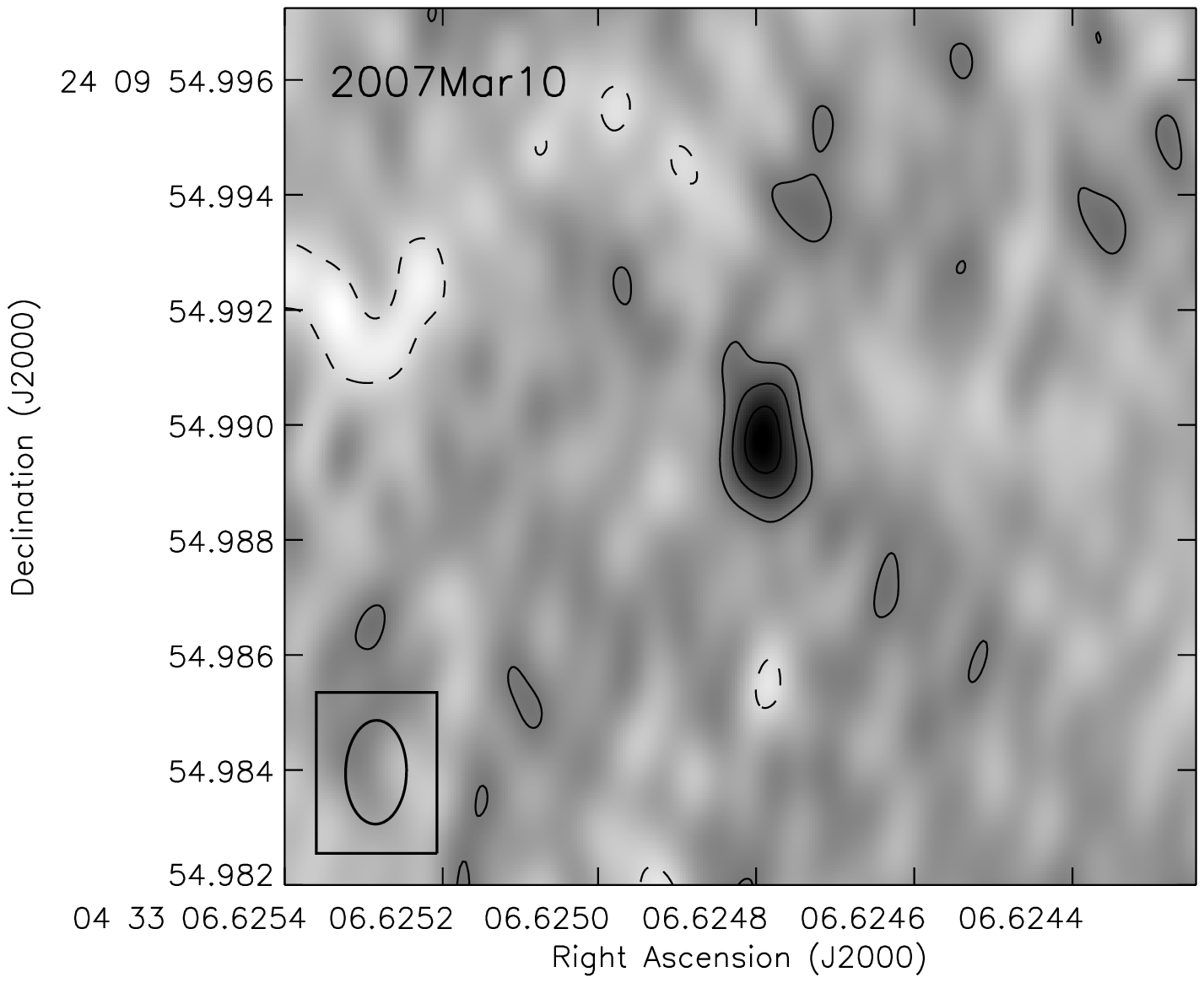}} 
   \scalebox{0.52}{\includegraphics{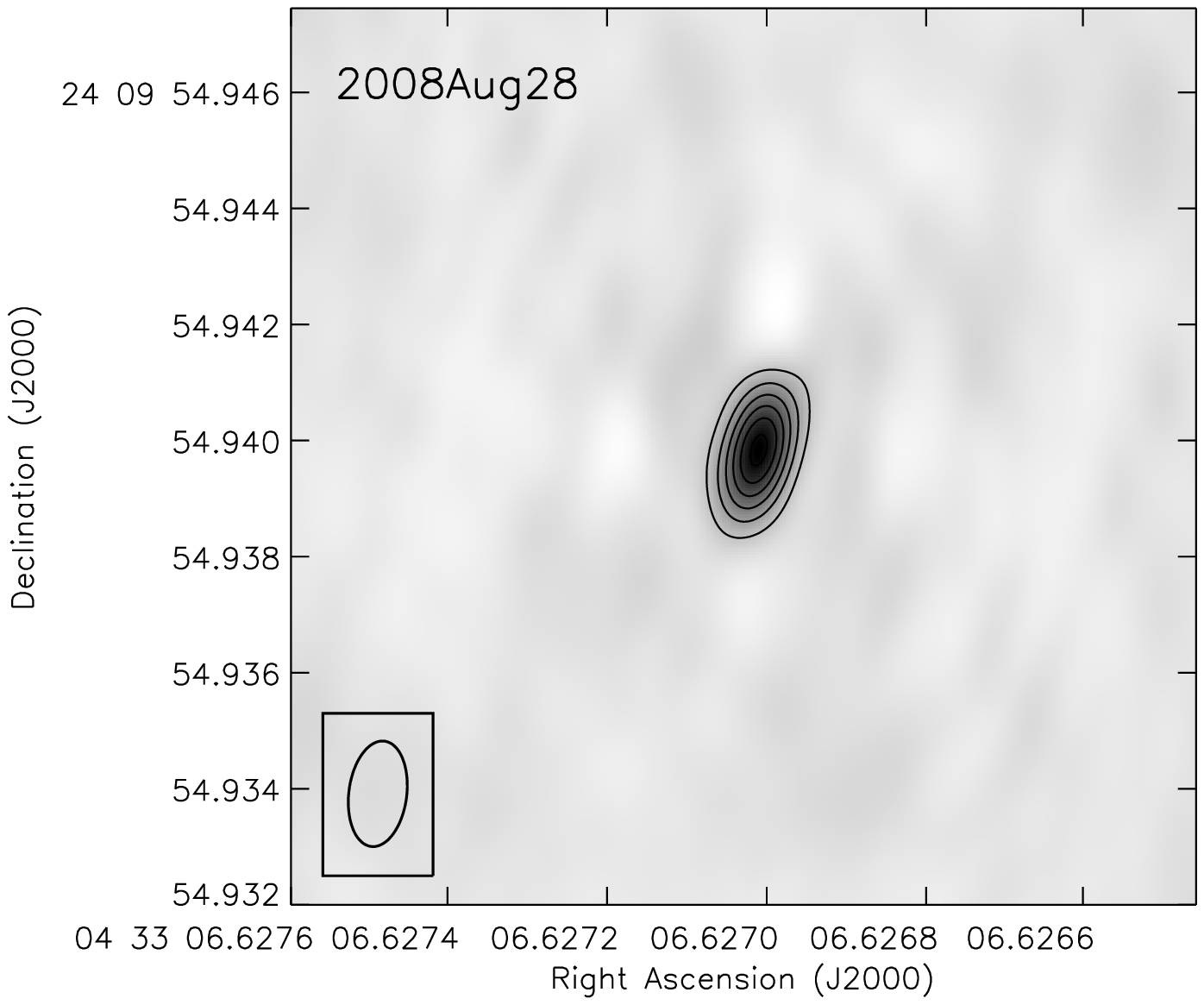}} \\
   \scalebox{0.8}{\includegraphics{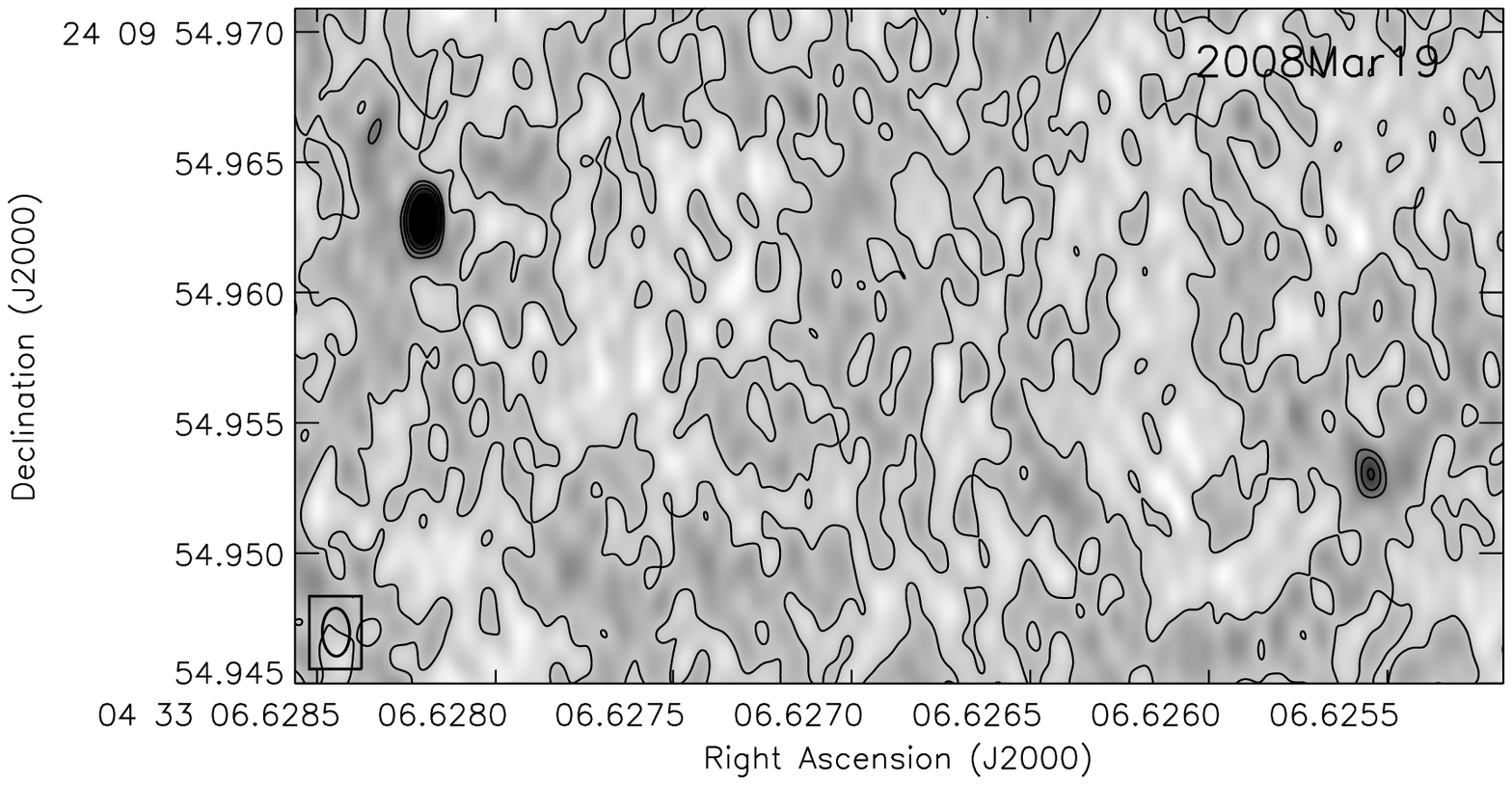}}
\end{center}
   \caption{VLBA maps of V807 Tau at 3.6 cm obtained on 2007 Mar 10, 2008 Mar 19, and 2008 Aug 28.  The contours correspond to intervals of the uncertainty in the integrated flux given by [-1, 1, 2, 3]$\times \sigma$, [0, 2, 3, 4, 6, 8, 10]$\times \sigma$, [5, 10, 15, 25, 30]$\times \sigma$, respectively for each epoch.  The inset shows the size and orientation of the synthesized beam.}
\label{fig.vlba_map}
\end{figure}

\clearpage

\begin{figure}
\begin{center}
   \scalebox{1.0}{\includegraphics{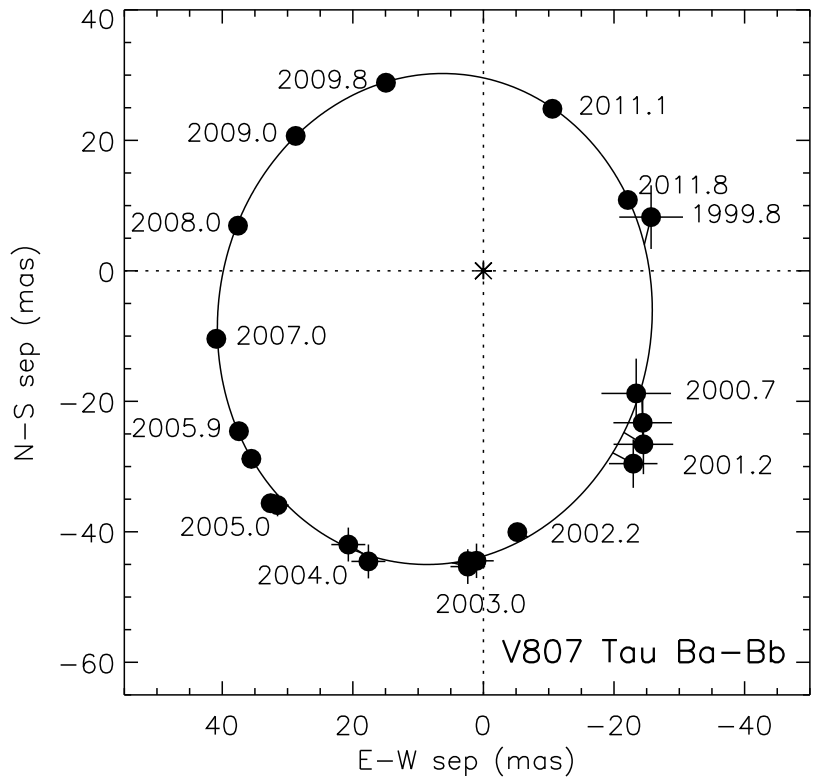}}
\end{center}
   \caption{Orbital motion of V807 Tau Bb relative to Ba
based on our adaptive optics and FGS observations.  Overplotted is the best fit orbit.}
\label{fig.orbit}
\end{figure}

\clearpage

\begin{figure}
\begin{center}
   \scalebox{1.0}{\includegraphics{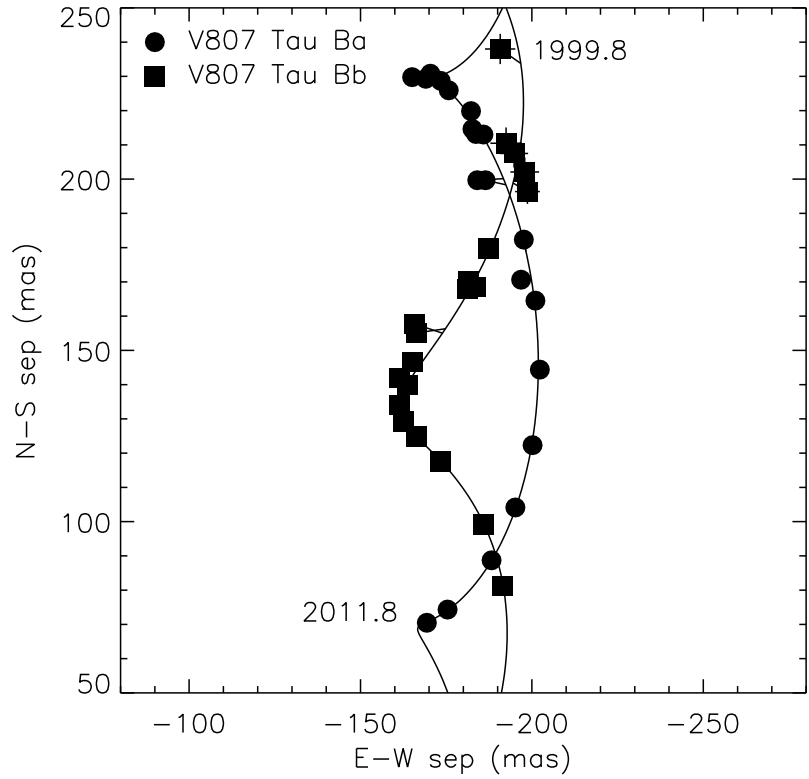}}
\end{center}
   \caption{Astrometric motion of V807 Tau Ba (circles) and Bb (squares) relative to the wide
component V807 Tau A (located at 0,0). The solid lines show
the astrometric motion of the close binary components with a
mass ratio of 0.843.}
\label{fig.astrom}
\end{figure}

\clearpage

\begin{figure}
\begin{center}
   \scalebox{1.0}{\includegraphics{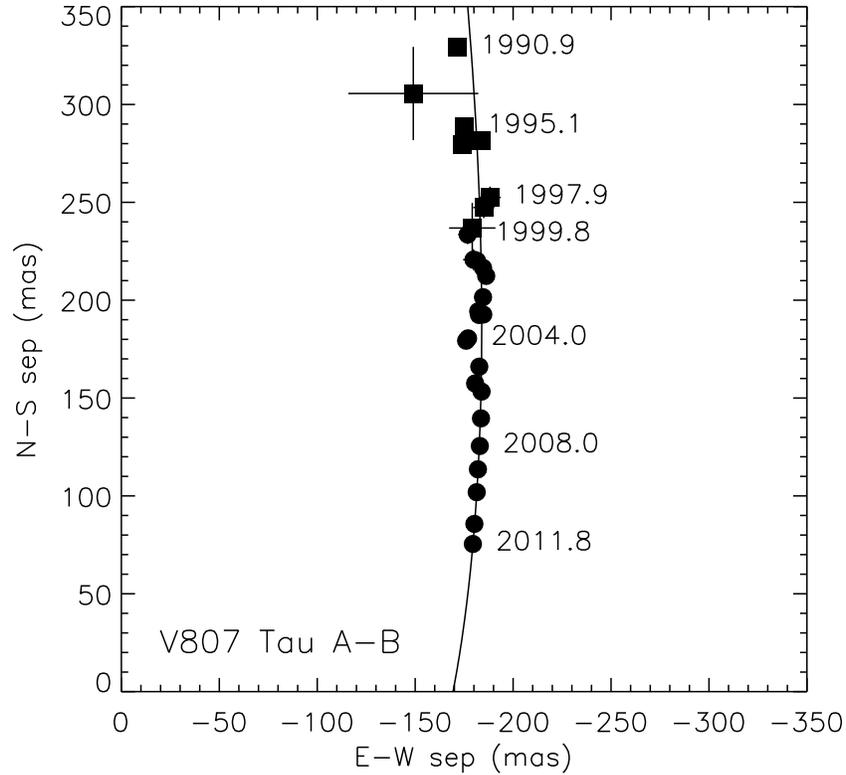}}
\end{center}
   \caption{Motion of the center of mass of V807 Tau Ba$-$Bb (filled circles) relative to V807 Tau A, computed from the derived mass ratio of 0.843 and the measured positions of the close components relative to V807 Tau A.  Overplotted is the representative orbit derived for the center of mass motion.  For comparison we plot the measured positions of the wide pair V807 Tau A$-$B (filled squares) reported in the literature \citep{ghez95,simon96,white01,hartigan03}.}
\label{fig.pos_cms}
\end{figure}

\clearpage

\begin{figure}
\begin{center}
   \scalebox{0.75}{\includegraphics{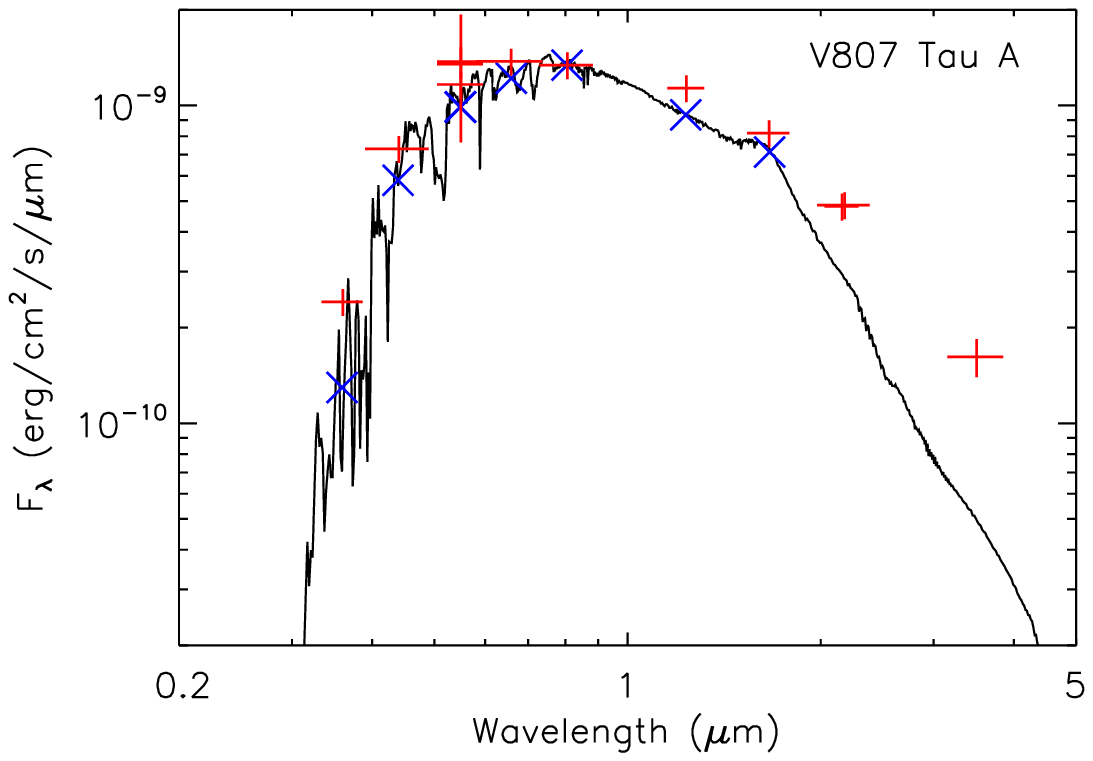}}
   \scalebox{0.75}{\includegraphics{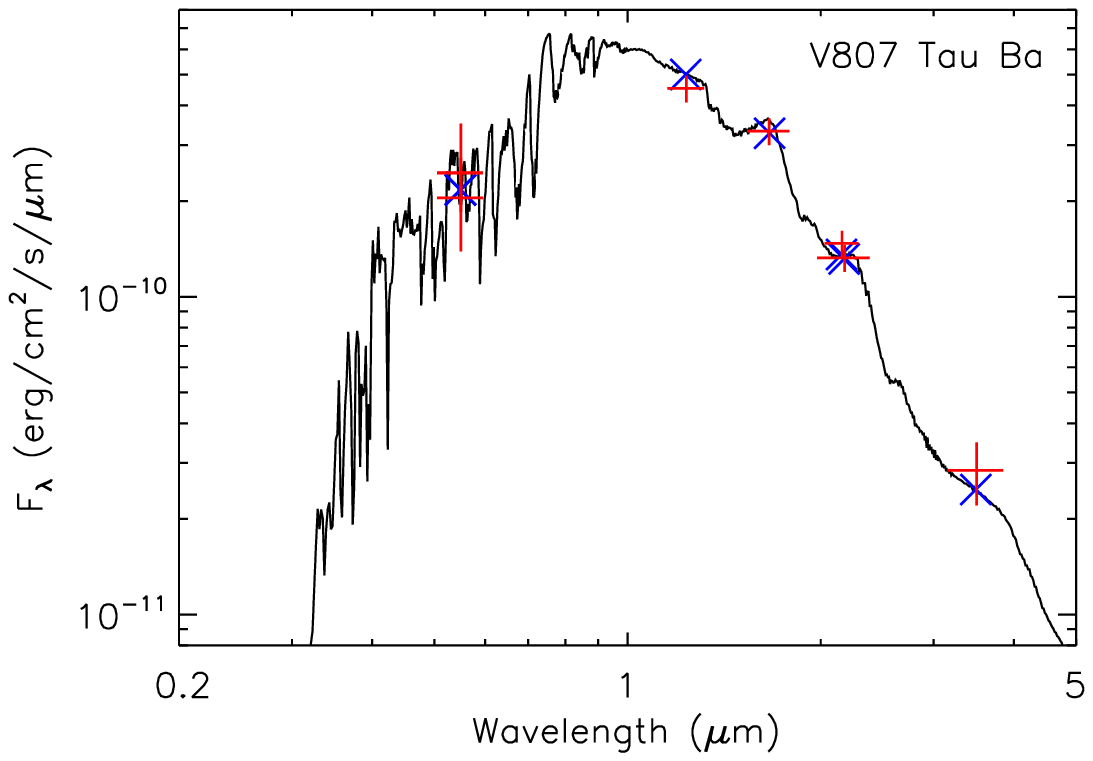}}
   \scalebox{0.75}{\includegraphics{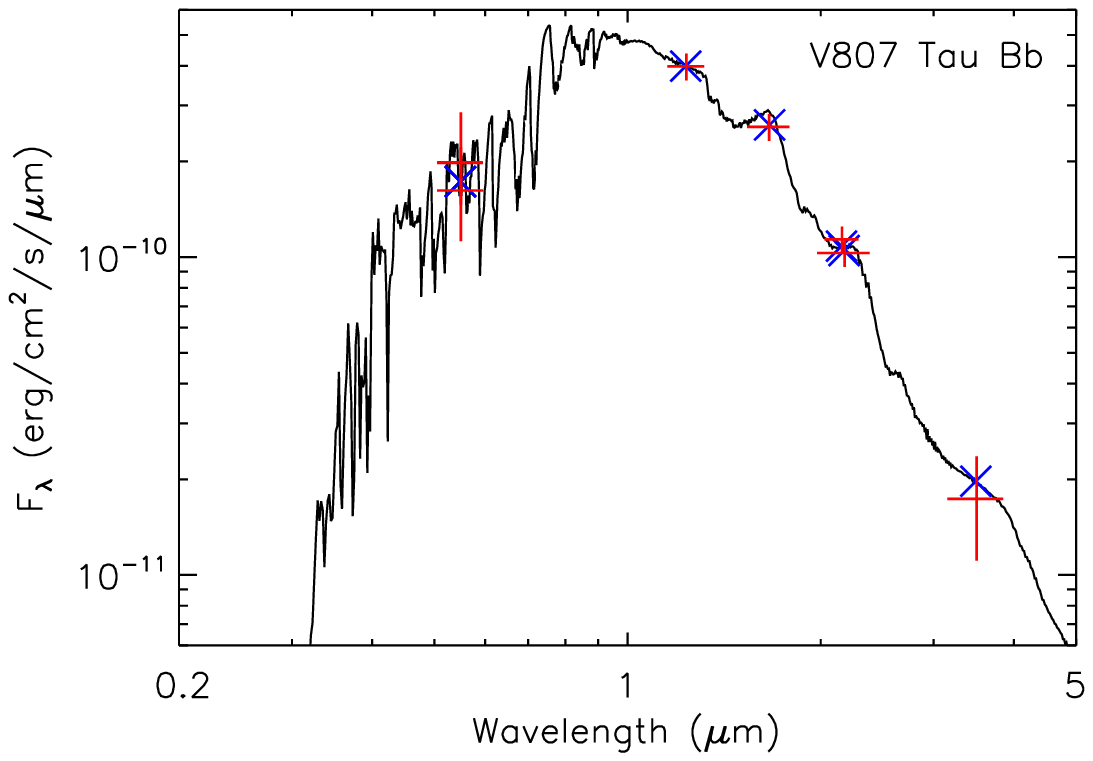}}
\end{center}
   \caption{Spectral energy distributions for V807 Tau A (top), V807 Tau Ba (middle), and V807 Tau Bb (bottom).  The best fitting Kurucz model is represented by the solid black line.  The average model flux across the given photometric band is shown by the blue crosses.  The red plus symbols show the measured photometric magnitudes; the vertical length gives the size of the measurement error in the magnitudes while the horizontal width represents the width of the photometric filter.}
\label{fig.sed}
\end{figure}

\clearpage

\begin{figure}
\begin{center}
   \scalebox{1.0}{\includegraphics{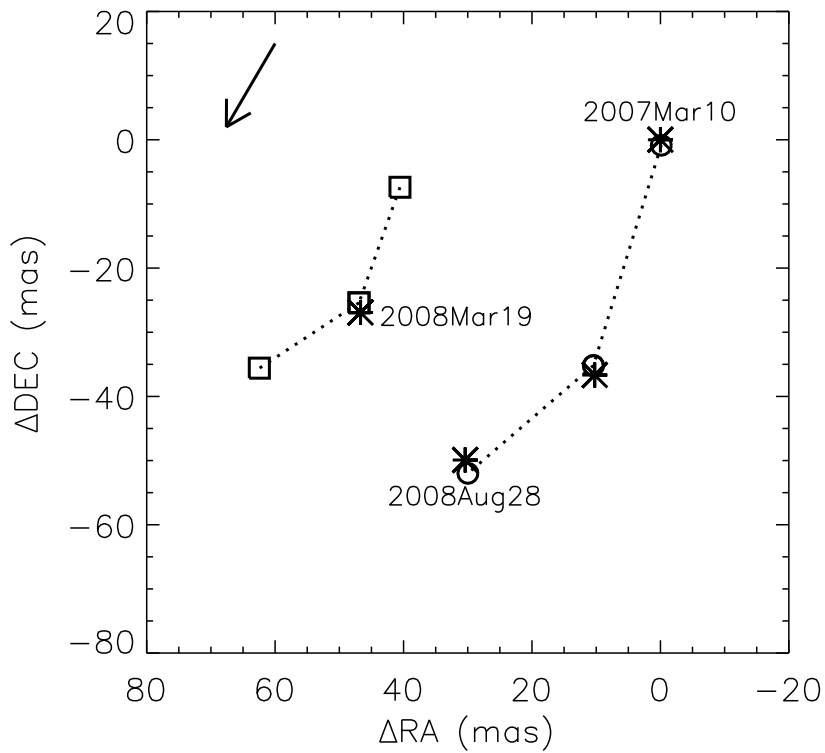}}
\end{center}
   \caption{Relative positions measured for the radio source
during the three VLBA epochs (asterisks). For reference, we
overplot the infrared positions of V807 Tau Ba and Bb
determined from the astrometric orbit during these epochs (circles and squares, respectively).
We applied our derived parallax of  $7.3 \pm 1.2$ mas and proper motion of $\mu_\alpha\cos{\delta} = 7.3 \pm 1.2$ mas\,yr$^{-1}$, $\mu_\delta = -13.0 \pm 1.4$ mas\,yr$^{-1}$ to the infrared positions.  The arrow indicates the direction and magnitude of the proper motion over the course of a year.}
\label{fig.vlba_motion}
\end{figure}

\clearpage

\begin{figure}
\begin{center}
   \scalebox{0.48}{\includegraphics{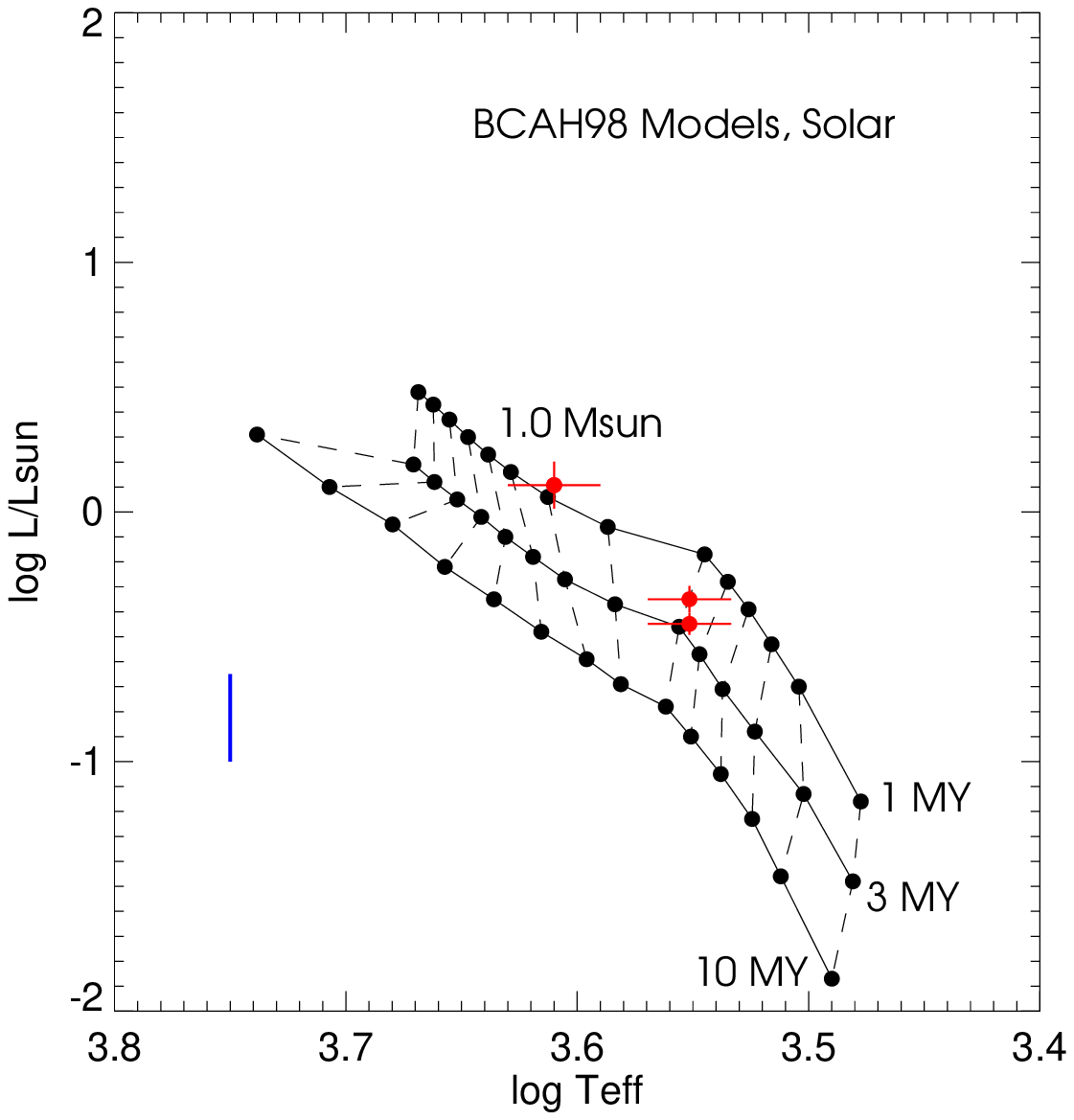}} 
   \scalebox{0.48}{\includegraphics{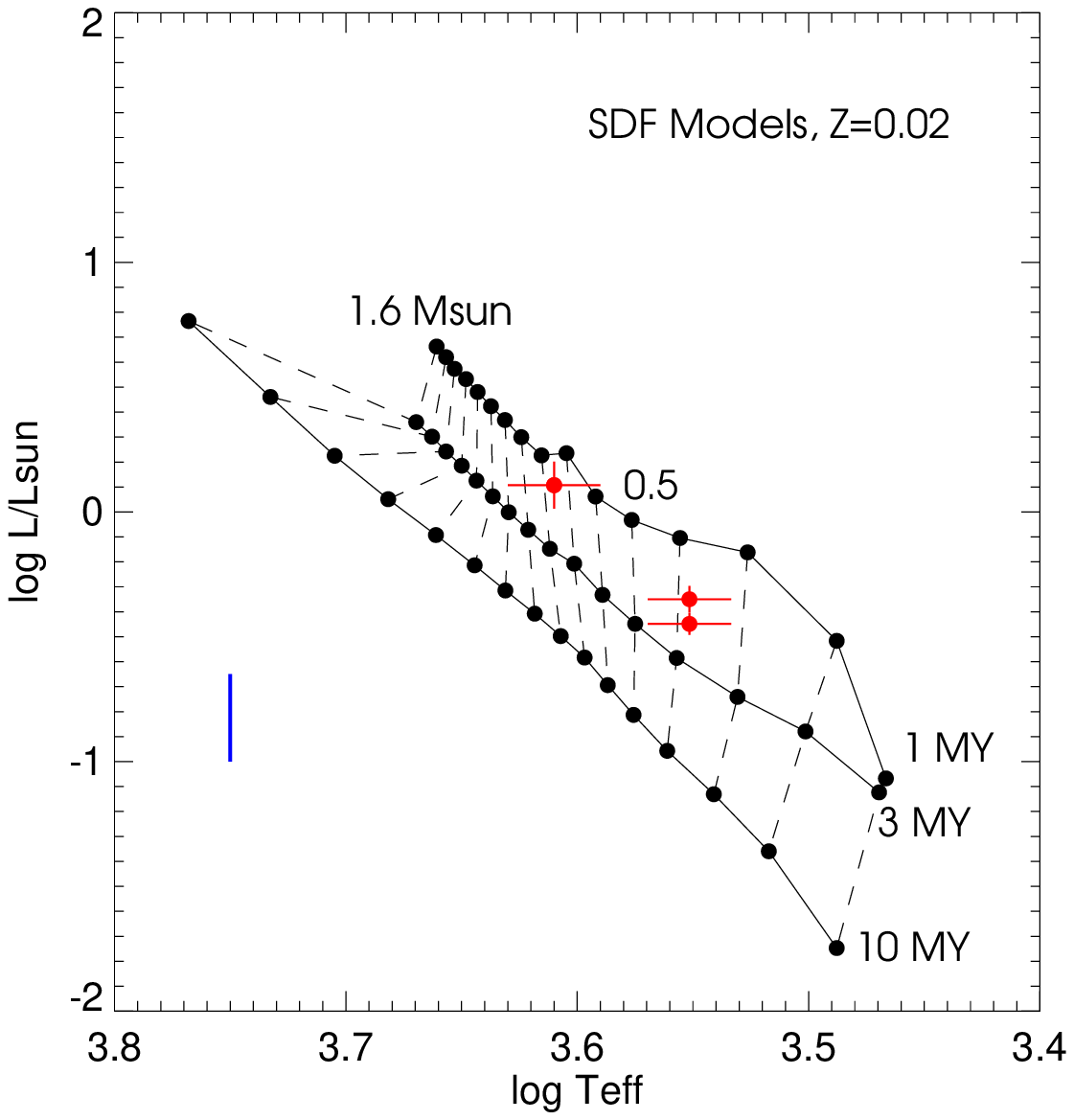}} 
   \scalebox{0.48}{\includegraphics{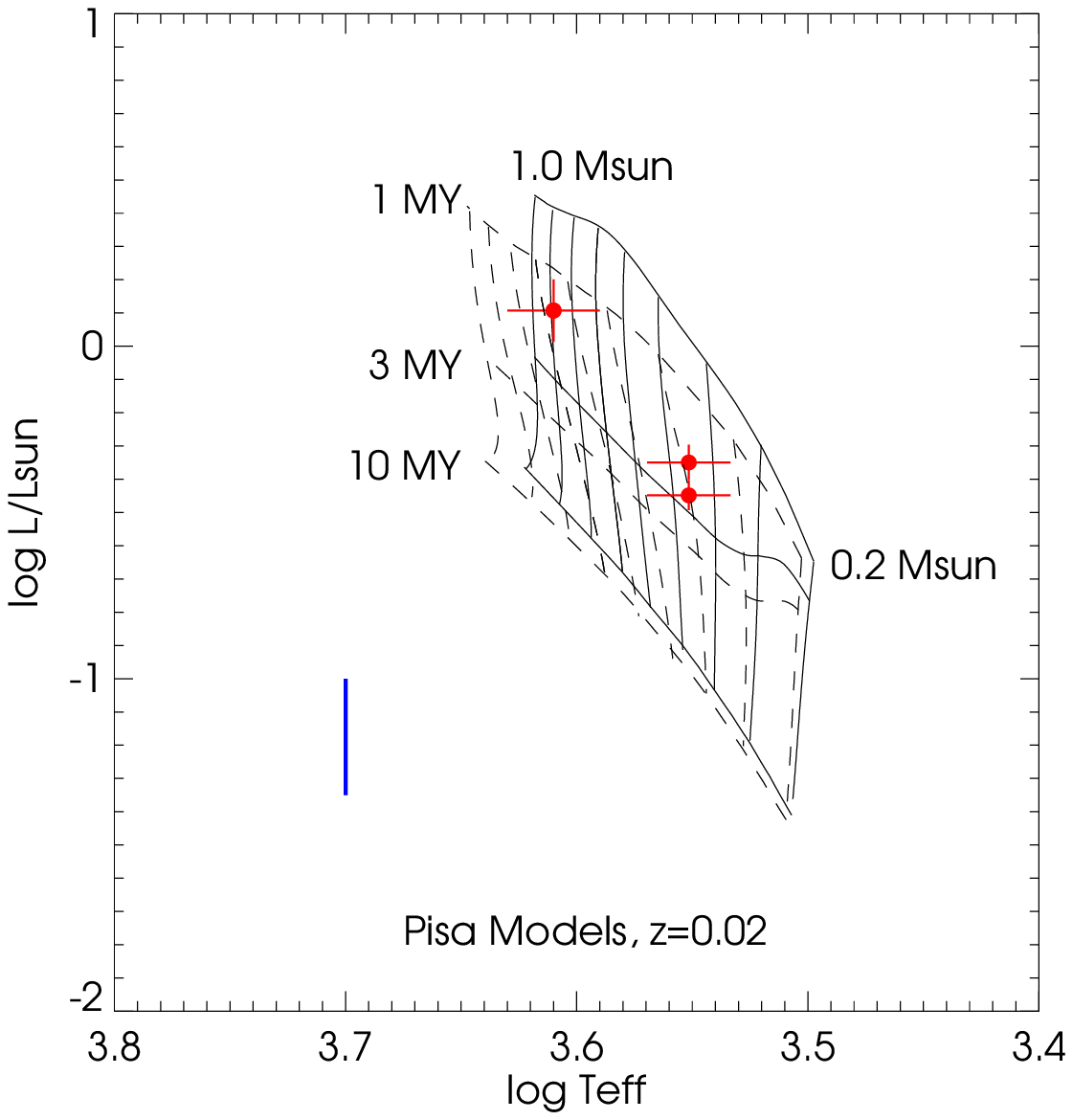}}
\end{center}
   \caption{HR diagrams showing the location of V807 Tau A, Ba, and Bb with respect to pre-main sequence evolutionary tracks.  {\it Left:} Evolutionary tracks of \citet[][BCAH98]{baraffe98} at solar metallicity for masses of 0.1 to 1.4 M$_\odot$.  For masses below 0.7 M$_\odot$, we used the tracks calculated with a mixing length parameter of 1.0; for larger masses we used the tracks with a mixing length of 1.9.  {\it Middle:} Evolutionary tracks of \citet[][SDF]{siess00} at $Z=0.02$ for masses of 0.1 to 1.6 M$_\odot$.  {\it Right:} Pisa  stellar evolutionary models \citep{tognelli11} for mixing length parameter of $\alpha = 1.20$ (solid lines) and $\alpha = 1.68$ (dashed lines) at a metallicity of $Z=0.02$.  The evolutionary tracks are plotted from 0.2 to 1.0 M$_\odot$ at 0.1 M$_\odot$ intervals.  In all of the panels, the isochrones are plotted at 1, 3, and 10 Myr.  The vertical blue bar represents the range of distances for stars located in the Taurus star forming region, $140 \pm 20$ pc.
}
\label{fig.hrd}
\end{figure}

%% If you use the table environment, please indicate horizontal rules using
%% \tableline, not \hline.
%% Do not put multiple tabular environments within a single table.
%% The optional \label should appear inside the \caption command.

%% If the table is more than one page long, the width of the table can vary
%% from page to page when the default \tablewidth is used, as below.  The
%% individual table widths for each page will be written to the log file; a
%% maximum tablewidth for the table can be computed from these values.
%% The \tablewidth argument can then be reset and the file reprocessed, so
%% that the table is of uniform width throughout. Try getting the widths
%% from the log file and changing the \tablewidth parameter to see how
%% adjusting this value affects table formatting.

%% The \dataset{} macro has also been applied to a few of the objects to
%% show how many observations can be tagged in a table.

%% You can append references to a table using the \tablerefs command.

%% Tables may also be prepared as separate files. See the accompanying
%% sample file table.tex for an example of an external table file.
%% To include an external file in your main document, use the \input
%% command. Uncomment the line below to include table.tex in this
%% sample file. (Note that you will need to comment out the \documentclass,
%% \begin{document}, and \end{document} commands from table.tex if you want
%% to include it in this document.)

%% \input{table}

%% The following command ends your manuscript. LaTeX will ignore any text
%% that appears after it.

\end{document}